\documentclass[aps,prl,twocolumn,amsmath,amssymb,showpacs,superscriptaddress,notitlepage,longbibliography]{revtex4-2}
\usepackage[colorlinks=true,linkcolor=blue,anchorcolor=red,citecolor=blue, urlcolor=blue]{hyperref}
\usepackage{bm}
\usepackage{graphicx}
\usepackage{xcolor}
\usepackage{wasysym}
\usepackage{makecell}
\usepackage{multirow}

\begin{document}
\title{Differential current noise as an identifier of Andreev bound states that induce nearly quantized conductance plateaus}

\author{Zhan Cao}
\email{Corresponding author.\\
caozhan@baqis.ac.cn}
\affiliation{Beijing Academy of Quantum Information Sciences, Beijing 100193, China}

\author{Gu Zhang}
\affiliation{Beijing Academy of Quantum Information Sciences, Beijing 100193, China}

\author{Hao Zhang}
\affiliation{State Key Laboratory of Low Dimensional Quantum Physics, Department of Physics, Tsinghua University, Beijing, 100084, China}
\affiliation{Beijing Academy of Quantum Information Sciences, Beijing 100193, China}
\affiliation{Frontier Science Center for Quantum Information, Beijing 100084, China}

\author{Ying-Xin Liang}
\affiliation{Beijing Academy of Quantum Information Sciences, Beijing 100193, China}

\author{Wan-Xiu He}
\affiliation{Beijing Computational Science Research Center, Beijing 100193, China}

\author{Ke He}
\affiliation{State Key Laboratory of Low Dimensional Quantum Physics, Department of Physics, Tsinghua University, Beijing, 100084, China}
\affiliation{Beijing Academy of Quantum Information Sciences, Beijing 100193, China}
\affiliation{Frontier Science Center for Quantum Information, Beijing 100084, China}
\affiliation{Hefei National Laboratory, Hefei 230088, China}

\author{Dong E. Liu}
\email{Corresponding author.\\
dongeliu@mail.tsinghua.edu.cn}
\affiliation{State Key Laboratory of Low Dimensional Quantum Physics, Department of Physics, Tsinghua University, Beijing, 100084, China}
\affiliation{Beijing Academy of Quantum Information Sciences, Beijing 100193, China}
\affiliation{Frontier Science Center for Quantum Information, Beijing 100084, China}
\affiliation{Hefei National Laboratory, Hefei 230088, China}

\begin{abstract}
Quantized conductance plateaus, a celebrated hallmark of Majorana bound states (MBSs) predicted a decade ago, have recently been observed with small deviations in iron-based superconductors and hybrid nanowires. Here, we demonstrate that nearly quantized conductance plateaus can also arise from trivial Andreev bound states (ABSs). To avoid ABS interruptions, we propose identifying ABS-induced quantized conductance plateaus by measuring the associated differential current noise $P$ versus bias voltage $V$. Specifically, for a quantized conductance plateau induced by one or multiple low-energy ABSs, the associated $P(V)$ curve exhibits a double-peak around zero bias, with the peak positions at $e|V|\approx 3k_B T$ (where $T$ is the temperature) and peak values larger than $2e^3/h$. These features greatly contrast those of an MBS or quasi-MBS, whose $P(V)$ curve displays a broad zero-bias dip and is consistently below $2e^3/h$. This protocol can be practically implemented in a variety of MBS candidate platforms using an electrode or STM tip as a probe.
\end{abstract}

\maketitle
Majorana bound states (MBSs)~\cite{read2000paired,kitaev2001unpaired} have attracted great attention due to their potential realization of topological quantum computation~\cite{kitaev2003fault,nayak2008non,sarma2015majorana}. They have been searched in many platforms~\cite{mourik2012signatures,deng2016majorana,fornieri2019evidence,ren2019topological,vaitiekenas2020flux,vaitiekenas2021zero,song2022large,cao2023recent,liu2018robust,wang2018evidence,machida2019zero,liu2020new,kong2021majorana,nadj2014observation,jeon2017distinguishing,xu2015experimental,sun2016majorana,flensberg2021engineered} by tunneling spectroscopy, inspired by the predicted MBS-induced quantized zero-bias conductance peak (ZBCP)~\cite{sengupta2001midgap,law2009majorana,flensberg2010tunneling}.
In turn, a discernible quantized ZBCP is believed as a significant benchmark of MBSs for implementing topological qubits~\cite{sau2020counting}. Thus far, most of the ZBCPs observed in Majorana candidate platforms are unconvincing due to possible interruption from false-positive signals generated by topologically trivial zero-energy Andreev bound states (ABSs)~\cite{liu2017andreev,pan2020physical,prada2020from}. Recently, this was tested by introducing a dissipative environment to suppress interrupting signals without sabotaging the robust tunneling feature of MBSs~\cite{liu2013proposed,liu2022universal,zhang2022suppressing,wang2022large,zhang2023theoretical,zhang2022situ}.

The quantized conductance plateau~\cite{wimmer2011quantum}, formed by a quantized ZBCP robust to the tuning of an experimental parameter, was firmly believed to be an MBS feature. In addition, quantized conductance plateaus can arise from robust quasi-MBSs (QMBSs)~\cite{moore2018quantized} that have MBS-like local characteristics but lack topological protection. Differentiating between MBSs and QMBSs, both characterized by Majorana-type wavefunctions, is not pursued in this work. Recently, nearly quantized conductance plateaus have been observed in the tunneling spectroscopy of iron-based superconductors~\cite{chen2019quantized,zhu2020nearly} by sweeping the STM tip-sample distance and of hybrid InAs-Al nanowire~\cite{wang2022plateau} by sweeping the magnetic field or gate voltages. However, a nearly quantized conductance plateau may also be caused by disorder-induced zero-energy ABSs, though suggested by numerical simulations~\cite{sarma2021disorder,pan2021quantized,zeng2022partially}. These results seriously challenge the validity of the conductance plateau hallmark of an MBS or robust QMBS (MBS/QMBS).

\begin{table}[t!]
\centering
\caption{Features of $P(V)$ curves for different bound states that induce nearly quantized conductance plateaus or ZBCPs.}\label{tb1}
\begin{tabular}{cccc}
\hline
\hline
Cases &\makecell[c]{A broad\\ zero-bias dip} &\makecell[c]{A double-peak\\ at $e|V|\approx 3k_BT$} &\makecell[c]{$P_{max}>$\\$2e^3/h$}\\
\hline
\makecell[c]{An MBS/QMBS} &\checked &$\times$ &$\times$\\
\hline
\makecell[c]{One or multiple ABSs}  &$\times$ &\checked &\checked\\
\hline
\makecell[c]{ An MBS/QMBS \\ and multiple ABSs}  &\checked &$\times$ &\checked\\
\hline
\hline
\end{tabular}
\end{table}

In this Letter, we theoretically study MBS/QMBS and zero-energy ABSs by combining their tunneling conductance and current noise. We demonstrate that a zero-energy ABS can also induce nearly quantized conductance plateaus in a sizable and continuous parameter space, which expands as temperature increases. We also show the possibility of nearly quantized ZBCPs when the probe is simultaneously coupled to multiple low-energy ABSs with or without an MBS/QMBS. We propose that the origin of a quantized conductance plateau or ZBCP can be identified by the associated differential current noise curve $P(V)$ (where $V$ is the bias voltage), which exhibits distinguishable features summarized in Table \ref{tb1}. These features are attributed to the strikingly distinct energy dependence of the eigenchannel transmissions in different cases. As both tunneling conductance and differential current noise can be accessed by the same probe with high precision~\cite{ferrier2016universality,bastiaans2018charge,bastiaans2021direct,thupakula2022coherent,liang2022low}, this protocol can be practically implemented in various MBS candidate platforms.

\begin{figure}[t!]
\centering
\includegraphics[width=0.9\columnwidth]{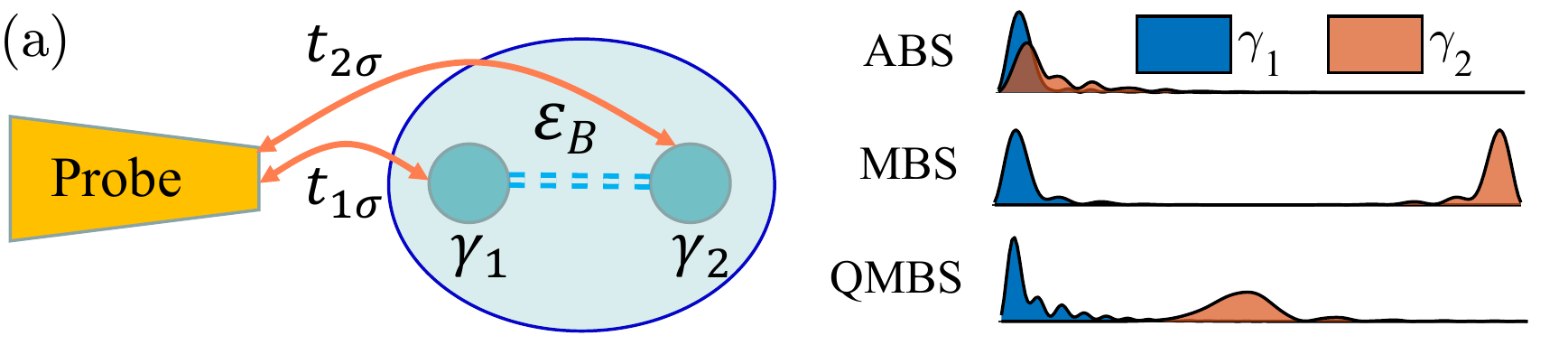}
\includegraphics[width=0.9\columnwidth]{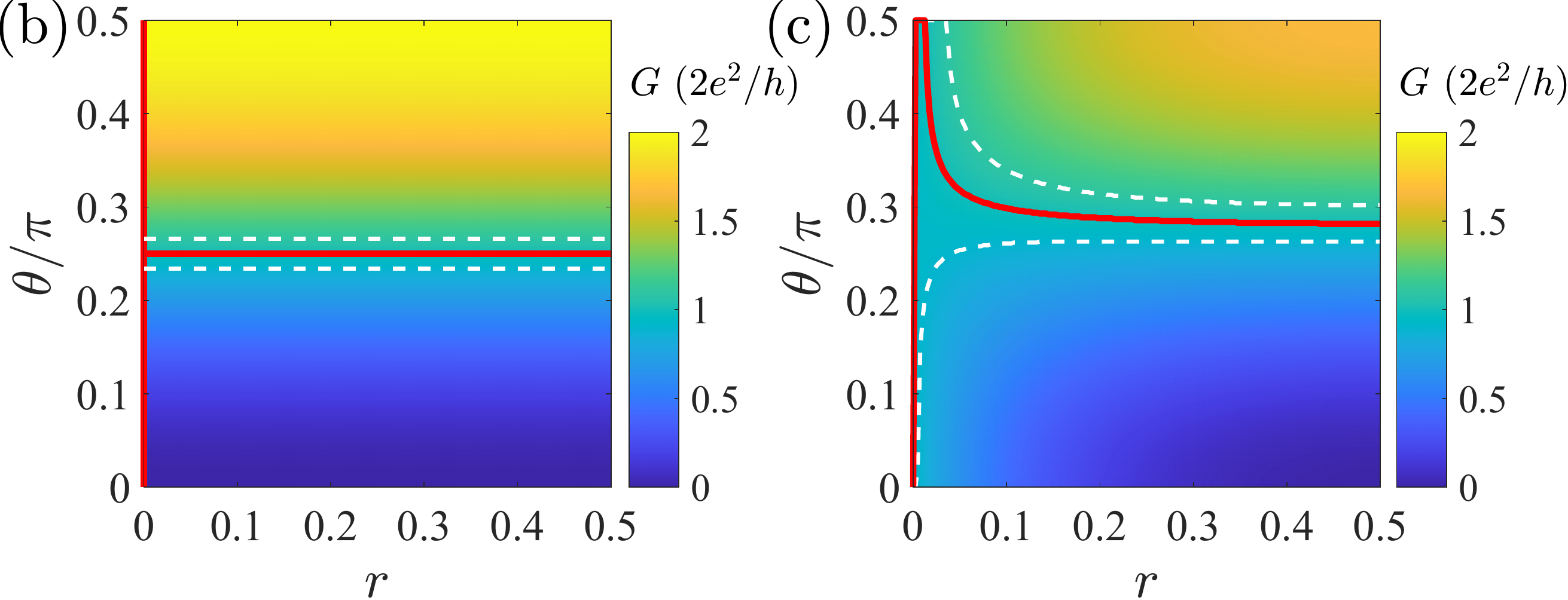}
\caption{(a) Left: schematic of a probe coupled to a superconducting bound state with energy $\varepsilon_B$. Physically, this bound state can be viewed as two hybridized Majoranas $\gamma_1$ and $\gamma_2$ that independently couple to the probe with spin-dependent amplitudes $t_{1,2\sigma}$. Right: typical profiles of 1D spatial wavefunctions of $\gamma_{1,2}$ for an ABS, MBS, and QMBS, respectively. They imply that $t_{1,2\sigma}\ne0$ for an ABS, while $t_{1\sigma}\ne0$ and $t_{2\sigma}=0$ for an MBS/QMBS. $t_{1,2\sigma}$ can be parametrized by real numbers $t_{1,2}$ and spin canting angles $\theta_{1,2}$ (see text), from which we define $\theta\equiv(\theta_1+\theta_2)/2$ and $r\equiv t_2^2/(t_1^2+t_2^2)$. $r=0$ ($r>0$) describes an MBS/QMBS (ABS). [(b) and (c)] The zero-bias conductance $G(0)$ for zero-energy states ($\varepsilon_B=0$) as functions of $r$ and $\theta$ are plotted at (b) zero temperature and (c) $k_B T=0.15~\Gamma$, respectively. The red solid curves highlight the lines with $G(0)=1$ and areas within the white dashed lines have $0.9\le G(0) \le 1.1$.}\label{Fig1}
\end{figure}

{\color{blue}\emph{Model and formalism}.}---We consider a local tunneling model schematically shown in Fig.~\ref{Fig1}(a) and described by the effective Hamiltonian $H=H_P+H_B+H_T$~\cite{flensberg2010tunneling,golub2011shot,vuik2019reproducing,danon2017conductance}.
Here $H_{P}=\sum_{k\sigma}\varepsilon_{k}c_{k\sigma}^{\dag}c_{k\sigma}$ describes a metallic probe (either an electrode or STM tip). $H_{B}=i\varepsilon_B \gamma_1\gamma_2$ describes a generic superconducting bound state as a composition of two Majorana components $\gamma_{1,2}$ with a hybridization energy $\varepsilon_B$. $H_{T}=\sum_{m k\sigma}(  t_{m\sigma}c_{k\sigma}^{\dag}-t_{m\sigma}^{\ast}c_{k\sigma})  \gamma_{m}$ describes the tunneling between the probe and $\gamma_{1,2}$. The spin-dependent hopping amplitudes $t_{m\sigma}$ ($m=1,2$) can be parametrized as~\cite{prada2017measuring} $t_{1\uparrow} =t_{1}\sin(\theta_1/2)$, $t_{1\downarrow}=-t_{1}\cos(\theta_1/2)$, $t_{2\uparrow}=-it_{2}\sin(\theta_2/2)$, and $t_{2\downarrow}=-it_{2}\cos(\theta_2/2)$, with $t_{1,2}$ being real numbers and $\theta_{1,2}$ the spin canting angles of the wave functions of $\gamma_{1,2}$ at the position nearest to the probe. In practice, $t_{1,2}$ and $\theta_{1,2}$ depend on experimentally tunable knobs, e.g., external magnetic field and gate voltages, and device details. As shown below, $\theta\equiv(\theta_1+\theta_2)/2$ is the relevant quantity for transport. For later convenience, we define the level broadening $\Gamma_{1,2}=2\pi\rho t_{1,2}^2$, with $\rho$ being the probe density of states, and assume $\Gamma_2\le \Gamma_1$ without loss of generality.

In Fig.~\ref{Fig1}(a), we illustrate the typical profiles of one-dimensional (1D) spatial wavefunctions of $\gamma_{1,2}$ for an ABS, MBS, and QMBS, respectively, that can potentially be probed. The two wave functions exhibit extensive overlap for an ABS, significant separation for an MBS, and partial separation for a QMBS~\cite{moore2018two,cao2019decays,vuik2019reproducing,stanescu2019robust,cao2022probing}. Consequently, both $\Gamma_1$ and $\Gamma_2$ are finite for an ABS, while $\Gamma_2 = 0$ for an MBS/QMBS. This implies that our local tunneling protocol cannot unequivocally distinguish MBS from robust QMBS. Nevertheless, our focus is to address whether nearly quantized conductance plateaus can be induced by trivial zero-energy ABSs and, if yes, how to experimentally identify them. For simplicity, we set $\varepsilon_B=0$ to focus on zero-energy bound states. Indeed, an ABS with zero energy can most closely resemble an MBS/QMBS. We further define $r=\Gamma_2/\Gamma$ (with $\Gamma=\Gamma_1+\Gamma_2$): $r = 0$ and $r>0$ for an MBS/QMBS and ABS, respectively.
Notably, earlier works study only limiting cases of the effective Hamiltonian under consideration: $r=0$ in Ref.~\onlinecite{flensberg2010tunneling}, $\theta=\pi/2$ in Refs.~\onlinecite{golub2011shot,vuik2019reproducing}, and $\sqrt{r(1-r)}\le\varepsilon_B/\Gamma$ in Ref.~\onlinecite{danon2017conductance}. By contrast, we visit the entire $(r,\theta)$ space that is experimentally accessible. The provided features of ABSs and MBS/QMBS are thus more complete and convincing.

With the scattering matrix theory~\cite{anantram1996current}, we derive the formulas of current $I$ and current noise $S$, refer to the Supplemental Material (SM)~\cite{Supp}. At zero temperature,
\begin{eqnarray}
&&I(V)=\frac{2e}{h}\sum_{s=\pm}\int^{eV}_0 d \varepsilon T_s(\varepsilon),\label{I0}\\
&&S(V)=\textrm{sgn}(V)\frac{8e^2}{h}\sum_{s=\pm}\int^{eV}_0 d \varepsilon T_s(\varepsilon)\big[1-T_s(\varepsilon)\big],\label{S0}
\end{eqnarray}
with
\begin{eqnarray}
T_{\pm}(  \varepsilon)=\Gamma^{2}\big[  \varepsilon^{2}p_{+}+p_{-}(  \varepsilon_{B}^{2}+\Gamma^{2}q_{-}/2)  \pm    \Pi\big]/\Lambda,\label{Tpm0}\\
\Pi=|\varepsilon|\big[\delta_{1}(  \delta_{2}\varepsilon^{2}+2p_{-}\varepsilon_{B}^{2})+\delta_{2}p_{-}(1/2-\delta_{1})  \Gamma^{2}\big]^{1/2},\label{Pi0}\\
\Lambda=(  \varepsilon^{2}-\varepsilon_{B}^{2})  ^{2}+(  q_{+}\varepsilon^{2}+\varepsilon_{B}^{2}q_{-})  \Gamma^{2}+\Gamma^{4}q_{-}^{2}/4.\label{Lambda0}
\end{eqnarray}
Here $V$ is the bias voltage, $\delta_{1}=1/2-2r(  1-r)  \cos^{2}\theta$, $\delta_{2}=(  1-2r)  ^{2}/2$, $p_{\pm}=(\delta_{1}\pm\delta_{2})/2  $, and $q_{\pm}=(  1\pm 2\delta_{2})/2$. $T_\pm(\varepsilon)$ are the transmissions of the two eigenchannels ``$\pm$'' of the scattering problem.
Notably, $T_-(\varepsilon)=0$ for a spin-polarized ABS (i.e., $\theta=0$) or an MBS/QMBS (i.e., $r=0$).

Theoretically, ZBCPs may also arise from multiple low-energy superconducting bound states~\cite{liu2012zero}. We examine this situation by extending $H_T$ and $H_B$ to include multiple bound states~\cite{Supp}. As the probe involves two spin-subbands, a maximum of two eigenchannels are available for transport. Therefore, Eqs.~\eqref{I0} and \eqref{S0} remain applicable and $T_\pm(\varepsilon)$ could be evaluated numerically. In this work, we mainly focus on the differential conductance $G(V)\equiv dI(V)/dV$ and differential current noise $P(V)\equiv |dS(V)/dV|$.

{\color{blue}\emph{Nearly quantized conductance plateau induced by a single zero-energy ABS}.}---At zero temperature, Eqs.~\eqref{I0}--\eqref{Lambda0} predict $G(0)=2e^2/h$ for $r=0$ and $G(0)=4e^2\sin^2\theta/h$ for $r>0$, which are shown in Fig.~\ref{Fig1}(b) as functions of $r$ and $\theta$. Along the red solid lines, $G(0)$ reaches the quantized value, and in the region between the two white dashed lines $0.9<G(0)<1.1$, showing a tolerance of $\pm 10\%$. This narrow parameter region agrees with previous simulations that a few percentages of disorder-induced zero-energy ABSs can lead to nearly quantized ZBCPs~\cite{pan2021quantized}. In Fig.~\ref{Fig1}(c), we show $G(0)$ at a finite temperature $k_BT=0.15~\Gamma$. The nearly quantized ZBCP region is larger than Fig.~\ref{Fig1}(b), indicating that false-positive ZBCPs are more likely to emerge for larger $k_B T/\Gamma$. The values of $k_B T/\Gamma$ in Refs.~\cite{zhu2020nearly} and \cite{wang2022plateau}, which reported nearly quantized conductance plateaus, are estimated as $0.27$ and $0.03$, respectively~\cite{Supp}. The parameter set $(r,\theta)$ depends on experimentally tunable knobs and device details. Figures \ref{Fig1}(b) and \ref{Fig1}(c) suggest that nearly quantized conductance plateaus induced by zero-energy ABSs could appear if the tuning of one or more knobs occasionally results in $(r,\theta)$ trajectories restricted between the white dashed lines. This highlights that a nearly quantized conductance plateau is not an exclusive hallmark of MBS/QMBS. Additionally, in the SM~\cite{Supp} we demonstrate that the Fano factor, calculated by $S$ and $I$, approaching $2$ ($1$) in the zero (large) bias limit at low temperatures is also not exclusive to MBS/QMBS, providing a possible explanation for a very recent experiment~\cite{ge2023single}.

\begin{figure}[t!]
\centering
\includegraphics[width=0.9\columnwidth]{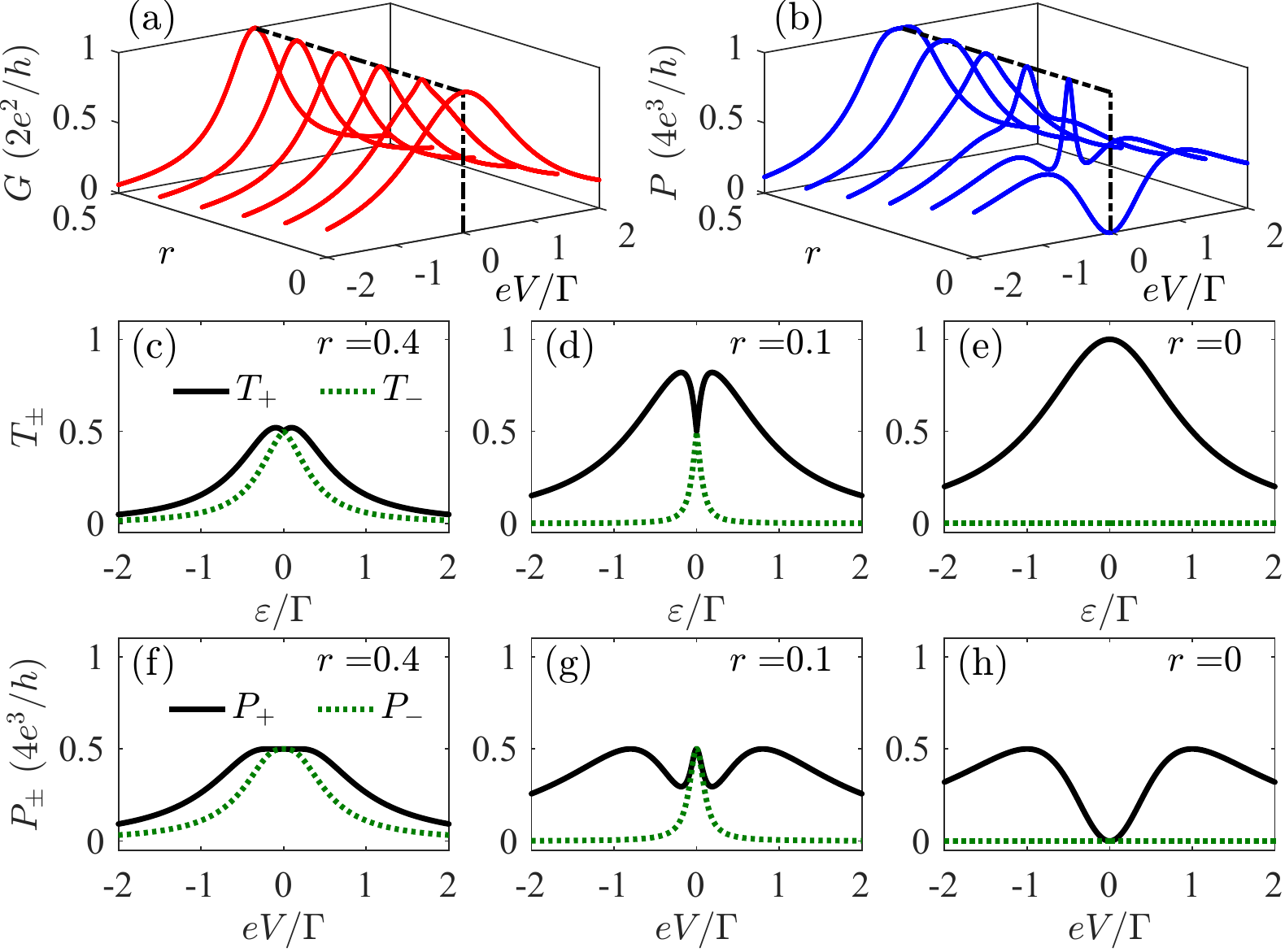}
\caption{[(a) and (b)] Zero-temperature differential conductance $G(V)$ and differential current noise $P(V)$ for zero-energy ABSs ($r> 0$) with $\theta=0.25\pi$ and MBS/QMBS ($r=0$). For $r=0.4$, $0.2$, and $0$, the [(c)--(e)] transmissions and [(f)--(h)] differential current noises associated with the two eigen channels ``$\pm$'' [see Eqs.~\eqref{I0}--\eqref{Tpm0}] are shown. 
}\label{Fig2}
\end{figure}

{\color{blue}\emph{Distinguish MBS/QMBS and ABSs induced quantized ZBCPs with differential current noise}.}---As shown in Fig.~\ref{Fig1}(b), zero-energy ABSs with $\theta=0.25\pi$ can induce a quantized zero-bias conductance, i.e., $G(0)=2e^2/h$. In Fig.~\ref{Fig2}(a), we further show the associated quantized ZBCPs in $G(V)$ curves for different $r$ values (from $0$ to $0.5$). The lineshapes of $G(V)$ for an MBS/QMBS ($r=0$) and ABS ($r>0$) are quite similar, making it hard to identify the origin of a quantized ZBCP. By contrast, the corresponding differential current noise $P(V)$ in Fig.~\ref{Fig2}(b) is quite distinct for an MBS/QMBS (a zero-bias dip) and ABS (a zero-bias peak). Furthermore, $P_{max}$, the maximum value of a $P(V)$ curve, is bounded by $2e^3/h$ for an MBS/QMBS, whereas it exceeds $2e^3/h$ for a zero-energy ABS. Such qualitative features in $P(V)$ can be easily discerned in experiments. For completeness, in the SM~\cite{Supp} we also present the correlations between $G(V)$ and $P(V)$ for ABSs with unquantized ZBCPs.

Now we analyze the features displayed in Figs.~\ref{Fig2}(a) and \ref{Fig2}(b).
As indicated by Eqs.~\eqref{I0} and \eqref{S0}, $G(V)$ and $P(V)$ are proportional to $\sum_{s=\pm}T_s(eV)$ and $\sum_{s=\pm}T_s(eV)[1-T_s(eV)]$, respectively.
The lineshapes of $T_\pm$, importantly, are different for an MBS/QMBS and ABS; see Figs.~\ref{Fig2}(c)--\ref{Fig2}(e).
Specifically, the channel ``$-$'' decouples [i.e., $T_-(\varepsilon)=0$] for an MBS/QMBS.
The MBS/QMBS induced $P(V)$ then follows the shape of $T_+ (1 - T_+)$: a broad dip at $T_+=1$ and two maximums of $2e^3/h$ at $T_+=0.5$ [Fig.~\ref{Fig2}(h)]. By contrast, for a generic ABS, the quantized ZBCP requires instead $T_+ (0) = T_- (0) = 0.5$ due to $T_\pm(0)=\sin^2\theta$. In this case, both channels are imperfect, leading to a peak of $P(0)=4e^3/h$ [with its components $P_\pm(V) $ shown in Figs.~\ref{Fig2}(f) and \ref{Fig2}(g)]. For a small $r$ value, the width of the zero-bias peak in $P(V)$ decreases as $ r \Gamma$~\cite{Supp}.

\begin{figure}[t!]
\centering
\includegraphics[width=0.9\columnwidth]{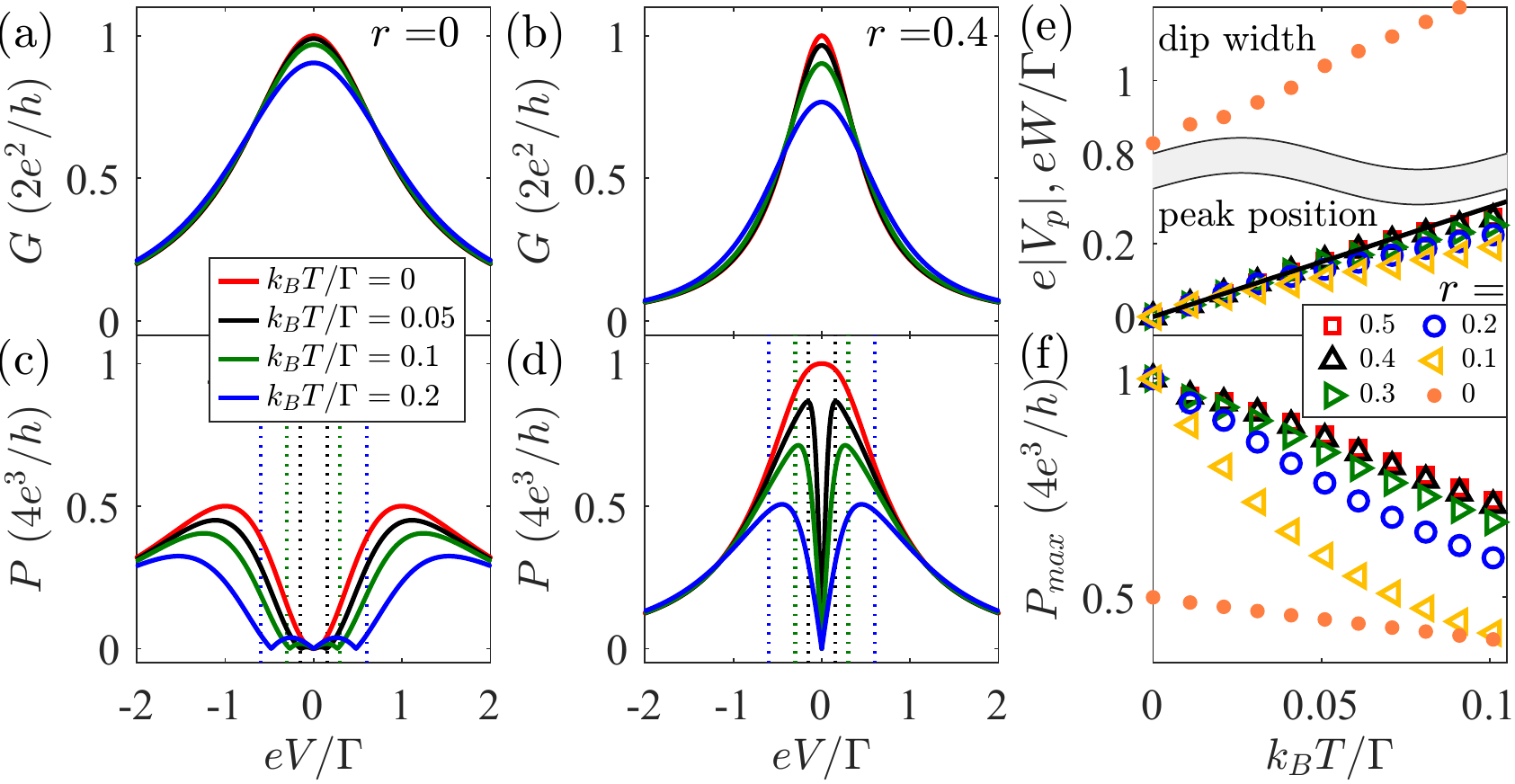}
\caption{[(a)--(d)] Temperature impact on the $G(V)$ and $P(V)$ curves of $r=0$ and $0.4$ in Fig.~\ref{Fig2}. Refer to the SM~\cite{Supp} for the results of the other $r$ values. The vertical dotted lines mark $e|V|= 3k_B T$. [(e) and (f)] Temperature dependence of the position $|V_p|$ of the double peak in $P(V)$ for ABSs ($r>0$), the width $W$ of the zero-bias dip in $P(V)$ for an MBS/QMBS ($r=0$), and the maximum $P_{max}$ of a $P(V)$ curve for ABS or MBS. The black solid line in (e) is a linear fit of $e\vert V_p\vert=3k_BT$.}\label{Fig3}
\end{figure}

Figure \ref{Fig3} shows the temperature impact on the $G(V)$ and $P(V)$ curves presented in Fig.~\ref{Fig2}. As temperature increases, the ZBCPs are suppressed below $2e^2/h$ by thermal broadening [Figs.~\ref{Fig3}(a) and \ref{Fig3}(b)]. Moreover, for an MBS/QMBS, the zero-bias dip in $P(V)$ is always broader than a constant $\sim\Gamma$ [Figs.~\ref{Fig3}(c) and \ref{Fig3}(e)]. On the other hand, for zero-energy ABSs, even if the zero-bias peak in $P(V)$ splits into a doublepeak, the interpeak distance ($\sim 6k_B T$) scales to zero as the temperature decreases [Figs.~\ref{Fig3}(d) and \ref{Fig3}(e)]. Additionally, as shown in Fig.~\ref{Fig3}(f), $P_{max}>2e^3/h$ is maintained for zero-energy ABSs when $k_BT/\Gamma\ll r$, while $P_{max}$ decreases from $2e^3/h$ with increasing temperature for an MBS/QMBS.

The temperature dependence of the peak position (dip width) of $P(V)$ for zero-energy ABSs (MBS/QMBS), as shown in Fig.~\ref{Fig3}(e), can be qualitatively understood as follows. In general, the current noise consists of shot noise, inherent to nonequilibrium charge transport even at $T=0$, and thermal noise, associated with equilibrium thermal fluctuations proportional to $f(\varepsilon-eV)[1-f(\varepsilon-eV)]$, where $f(\varepsilon-eV)$ is the Fermi distribution and the width of its derivative is roughly $3k_B T$. Consequently, at $T=0$, thermal noise is absent, and shot noise dominates. At $T>0$ and in the low-bias region $e|V|< 3k_B T$, thermal noise emerges and prevails over shot noise. 
% In the zero-bias limit, the derivative of $f(\varepsilon-eV)[1-f(\varepsilon-eV)]$ with respect to $\varepsilon$ is an odd function, resulting in $P(0)=0$ in Figs.~\ref{Fig3}(c) and \ref{Fig3}(d).
When the bias increases to the threshold $e|V|\approx 3k_B T$, above which shot noise dominates the current noise, the characteristic peak in $P(V)$ predicted for zero-energy ABSs at $T=0$ reappears. Due to the same reason, the dip width in $P(V)$ for an MBS/QMBS also increases by approximately $3k_B T$. 

\begin{figure}[t!]
\centering
\includegraphics[width=0.9\columnwidth]{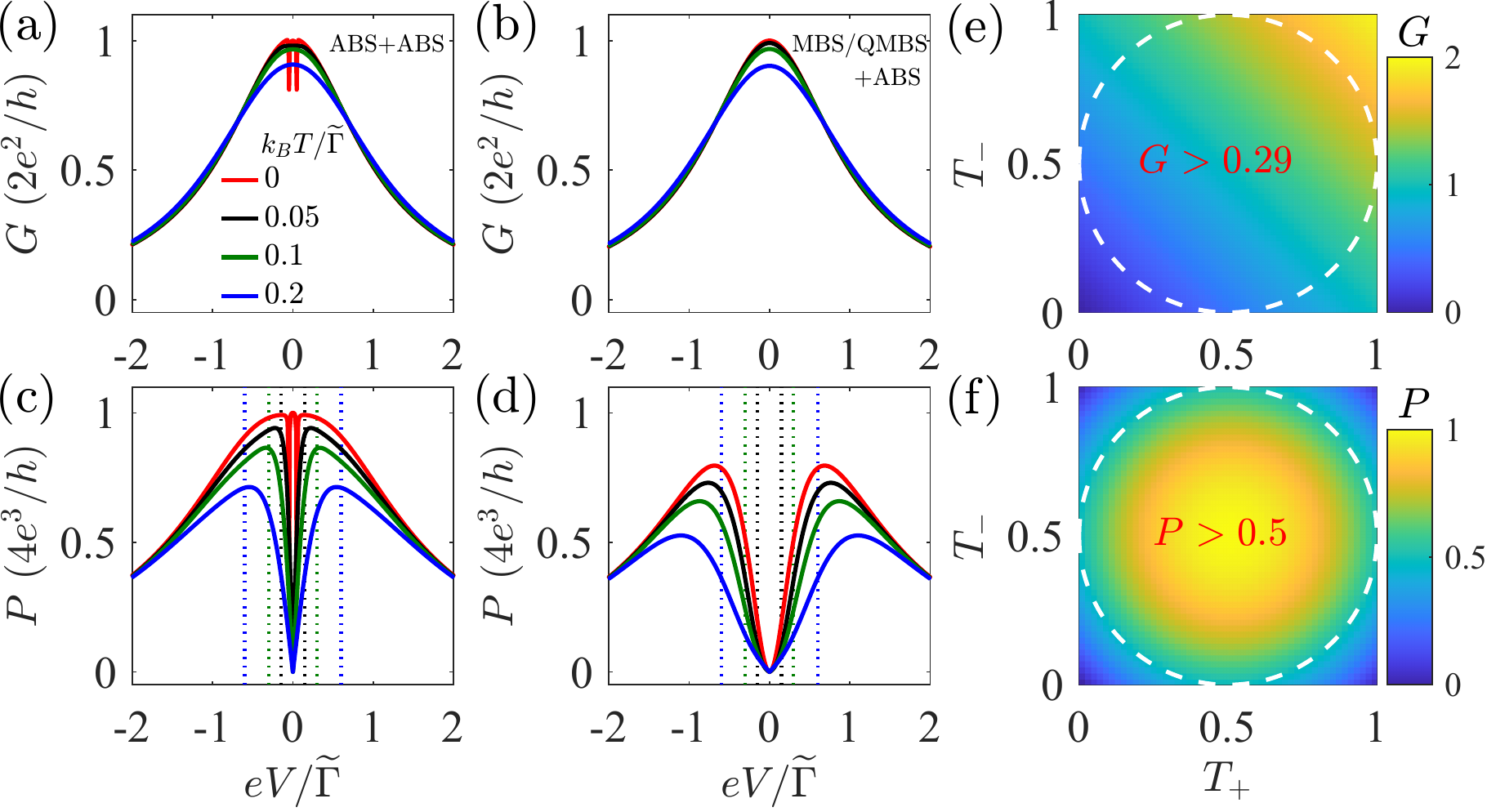}
\caption{Examples of $G(V)$ and $P(V)$ curves for a probe simultaneously coupled to [(a) and (c)] two ABSs with energies of $0$ and $0.1\widetilde\Gamma$, respectively, or to [(b) and (d)] an MBS/QMBS and a zero-energy ABS. The bias voltage is rescaled by $\widetilde\Gamma$ such that $G=e^2/h$ for $e|V|/\widetilde\Gamma=1$ and $T=0$. The vertical dotted lines mark $e|V|= 3k_B T$. Refer to the SM~\cite{Supp} for the calculation parameters. [(e) and (f)] Zero-temperature $G$ and $P$ as functions of $T_{\pm}$, calculated by Eqs.~\eqref{I0} and \eqref{S0}. The lower limits of $G$ and $P$ within the dashed circle are indicated.}\label{Fig4}
\end{figure}

Essentially, a ZBCP can potentially be induced by multiple low-energy bound states. For this scenario, we explore the correlations between $G(V)$ and $P(V)$ for a probe simultaneously coupled to a pair of low-energy ABSs or to a combination of an MBS/QMBS and a zero-energy ABS. Exemplary numerical results are shown in Fig.~\ref{Fig4}, which demonstrate that nearly quantized ZBCPs can occur in both cases [Figs.~\ref{Fig4}(a) and \ref{Fig4}(b)]. The associated $P(V)$ curves exhibit a double peak at $e|V|\approx 3k_B T$ [Fig.~\ref{Fig4}(c)] and a broad zero-bias dip [Fig.~\ref{Fig4}(d)], respectively. Both eigenchannels ``$\pm$'' contribute to $G$ and $P$, as shown in the SM~\cite{Supp}. The profiles of the $G(V)$ and $P(V)$ curves for two bound states in Figs.~\ref{Fig4}(a) and \ref{Fig4}(c) [Figs.~\ref{Fig4}(b) and \ref{Fig4}(d)] can be observed when a probe is coupled to multiple ABSs (an MBS/QMBS and multiple ABSs). Indeed, even with multiple bound states, there are at most two eigenchannels with transmissions $T_\pm$, as only two spin subbands are available in the probe. Notably, the low-temperature $P(V)$ curves in Figs.~\ref{Fig4}(c) and \ref{Fig4}(d) all exhibit $P_{max}>2e^3/h$, which is actually a universal feature when the probe is coupled to one or more ABSs. Specifically, as indicated by Figs.~\ref{Fig4}(e) and \ref{Fig4}(f), which are calculated by Eqs.~\eqref{I0} and \eqref{S0}, as long as both channels contribute to the transport with the condition that the maximum of a $G(V)$ curve exceeds $0.29 \times 2e^2/h$, we invariably find $P_{max}> 2e^3/h$.

\begin{figure}[t!]
\centering
\includegraphics[width=0.9\columnwidth]{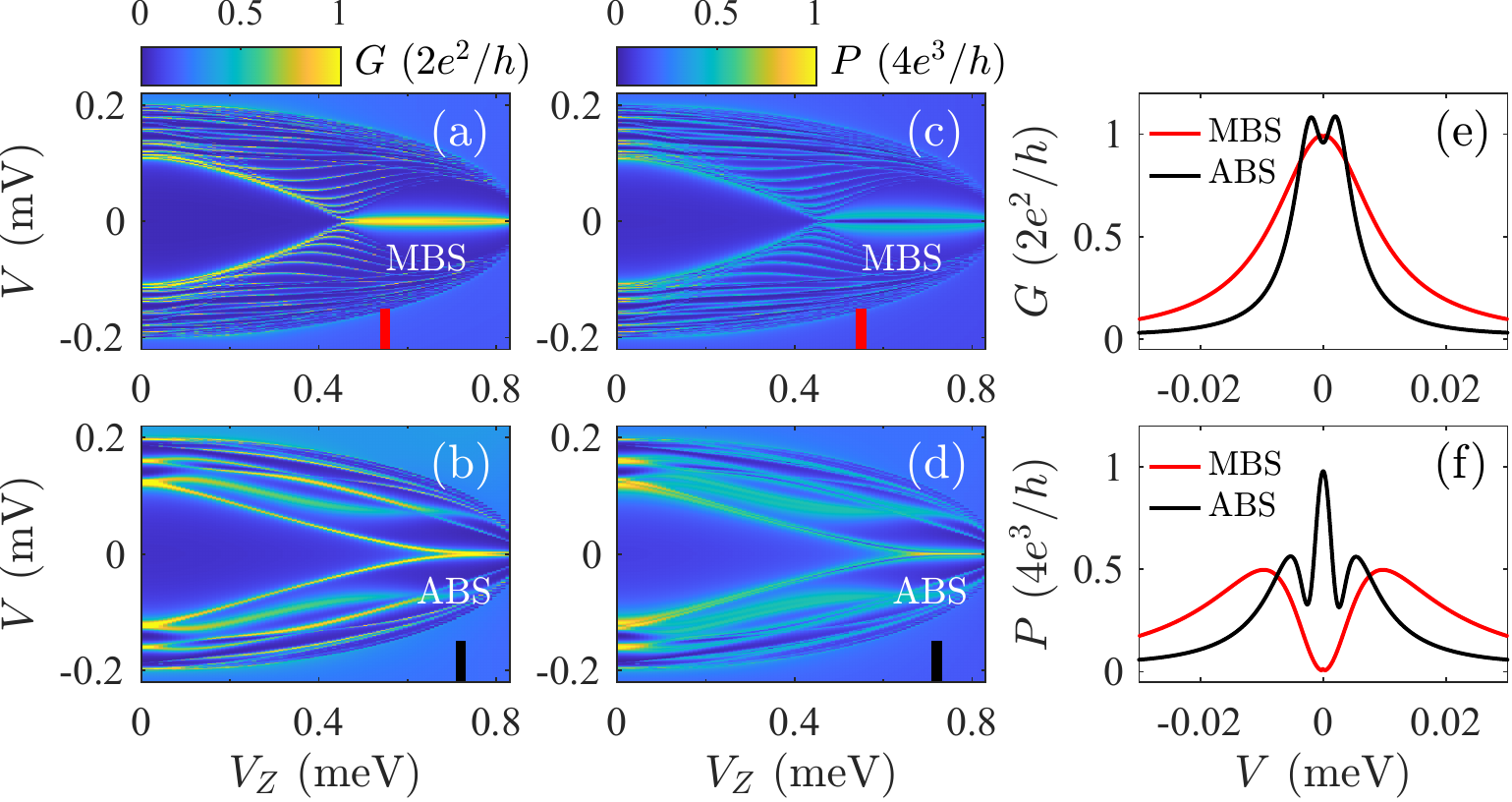}
\caption{[(a)--(d)] Zero-temperature simulations of $G$ and $P$ versus the bias voltage $V$ and the Zeeman energy $V_Z$ for a [(a) and (c)] clean and [(b) and (d)] disordered hybrid nanowire. [(e) and (f)] Line cuts of (a)--(d) at the $V_Z$ marked by the colored bars. Refer to the SM~\cite{Supp} for simulation parameters.}\label{Fig_simulation}
\end{figure}

{\color{blue}\emph{Simulations of hybrid nanowires}.}---The analyses above are based on the effective models of isolated bound states. To validate the reliability, we perform numerical simulations to obtain the $G$ and $P$ of MBSs and near-zero-energy ABSs in hybrid nanowires, considering several physical effects in experiments~\cite{liu2017andreev,pan2020physical,Supp}. In Fig.~\ref{Fig_simulation}, we compare the maps of $G$ and $P$, as functions of the bias voltage $V$ and the Zeeman energy $V_Z$, for a clean and disordered single-subband nanowire, respectively. Clearly, a nearly quantized conductance plateau persists over a sizable range of $V_Z$ for both nanowires [Figs.~\ref{Fig_simulation}(a) and \ref{Fig_simulation}(b)]. The local density of states shown in the SM~\cite{Supp} reveals that the two plateaus are associated with an MBS and near-zero-energy ABS, respectively. The $P$ maps of an MBS and ABS are qualitatively distinct [Figs.~\ref{Fig_simulation}(c) and \ref{Fig_simulation}(d)], which is evident from the line cuts in Figs.~\ref{Fig_simulation}(e) and \ref{Fig_simulation}(f): while both $G(V)$ curves display a nearly quantized ZBCP, the $P(V)$ curve for an MBS exhibits a broad zero-bias dip and $P_{max}\le2e^3/h$, whereas that for an ABS shows a zero-bias peak and $P_{max}>2e^3/h$. These $G(V)$ and $P(V)$ correlations qualitatively agree with those in Figs.~\ref{Fig2}(a) and \ref{Fig2}(b) predicted by the effective model.

Further simulations are presented in the SM~\cite{Supp}. We simulate four additional scenarios exhibiting nearly quantized conductance plateaus: (i) an MBS in a dissipated single-subband nanowire, (ii) a near-zero-energy ABS in a dissipated single-subband nanowire, (iii) a near-zero-energy ABS in the top subband of a two-subband nanowire, and (iv) two low-energy ABSs in different subbands of a two-subband nanowire. It is noteworthy that the results of all these scenarios consistently align with our predictions. For a comprehensive summary of our effective model calculations and numerical simulations, we refer the reader to Tables S3 and S4 in the SM~\cite{Supp}.

{\color{blue}\emph{Summary and Discussion}.}---We have demonstrated that the recently observed nearly quantized conductance plateaus~\cite{chen2019quantized,zhu2020nearly,wang2022plateau} can potentially arise from topologically trivial zero-energy or near-zero-energy ABSs. Moreover, we have proposed measuring the associated differential current noise to identify this scenario. While our protocol cannot distinguish between MBSs and robust QMBSs, it facilitates achieving a crucial near-term objective: assessing whether a given material system attains satisfactory quality to rule out trivial zero-energy ABSs. This is currently a major concern in hybrid nanowires~\cite{sarma2021disorder,pan2021quantized,ahn2021estimating,hess2021local,cao2022numerical,hess2023trivial,sarma2023spectral} and superconducting vortices~\cite{de2023near}. With improved material quality, the identification of MBSs could be accomplished through complementary evidence from different combined measurement protocols using sophisticated devices, such as the combination of local and nonlocal tunneling conductance in a three-terminal device~\cite{menard2020conductance,puglia2021closing,wang2022parametric,aghaee2023inas,poschl2022nonlocal,banerjee2023local}, local tunneling and Coulomb conductance in a two-terminal island device~\cite{valentini2022majorana}, or tunneling conductance and differential current noise in a quantum dot based three-terminal device~\cite{sergey2022revealing}.

{\color{blue}\emph{Acknowledgements}.}---We appreciate the anonymous referees' suggestion to consider the scenario involving multiple bound states. This work was supported by the National Natural Science Foundation of China (Grants No.~12004040 and No.~11974198), the Innovation Program for Quantum Science and Technology (Grant No.~2021ZD0302400), the National Natural Science Foundation of China (Grants No.~92065206 and No.~12374158), the National Key Research and Development Program of China (Grant No.~2017YFA0303303), and the Tsinghua University Initiative Scientific Research Program.

%\bibliographystyle{apsrev4-1-title}
%\bibliography{refs-Majorana}


\end{document}

\title{Supplemental Material\\
for\\
``Differential current noise as an identifier of Andreev bound states that induce nearly quantized conductance plateaus''}

\author{Zhan Cao}
\email{Corresponding author.\\
caozhan@baqis.ac.cn}
\affiliation{Beijing Academy of Quantum Information Sciences, Beijing 100193, China}

\author{Gu Zhang}
\affiliation{Beijing Academy of Quantum Information Sciences, Beijing 100193, China}

\author{Hao Zhang}
\affiliation{State Key Laboratory of Low Dimensional Quantum Physics, Department of Physics, Tsinghua University, Beijing, 100084, China}
\affiliation{Beijing Academy of Quantum Information Sciences, Beijing 100193, China}
\affiliation{Frontier Science Center for Quantum Information, Beijing 100084, China}

\author{Ying-Xin Liang}
\affiliation{Beijing Academy of Quantum Information Sciences, Beijing 100193, China}

\author{Wan-Xiu He}
\affiliation{Beijing Computational Science Research Center, Beijing 100193, China}

\author{Ke He}
\affiliation{State Key Laboratory of Low Dimensional Quantum Physics, Department of Physics, Tsinghua University, Beijing, 100084, China}
\affiliation{Beijing Academy of Quantum Information Sciences, Beijing 100193, China}
\affiliation{Frontier Science Center for Quantum Information, Beijing 100084, China}
\affiliation{Hefei National Laboratory, Hefei 230088, China}

\author{Dong E. Liu}
\email{Corresponding author.\\
dongeliu@mail.tsinghua.edu.cn}
\affiliation{State Key Laboratory of Low Dimensional Quantum Physics, Department of Physics, Tsinghua University, Beijing, 100084, China}
\affiliation{Beijing Academy of Quantum Information Sciences, Beijing 100193, China}
\affiliation{Frontier Science Center for Quantum Information, Beijing 100084, China}
\affiliation{Hefei National Laboratory, Hefei 230088, China}

%\date{\today}

\maketitle
\tableofcontents

\section{Derivation of the formulas of current and current noise}\label{sec1}
In this section, we present a detailed derivation of the formulas of current and current noise following the scattering matrix theory developed for noninteracting normal metal-superconductor hybrid systems~\cite{anantram1996current}. Applying the general formalism in Ref.~\onlinecite{anantram1996current} to the effective Hamiltonian considered in the main text, the current and current noise are
\begin{eqnarray}
I(V)&=&\frac{e}{h}\sum_{\alpha,\beta\in e,h}\textrm{sgn}(  \alpha)\int_{0}^{\infty}d\varepsilon \textrm{Tr}\big[  A^{\beta\beta}(  \alpha,\varepsilon)  \big]  f_{\beta}(  \varepsilon),\label{I}\\
S(V)&=&\frac{2e^{2}}{h}\sum_{\alpha,\beta,\gamma,\delta\in e,h}\textrm{sgn}(  \alpha)  \textrm{sgn}(  \beta)  \int_{0}^{\infty} d\varepsilon  \textrm{Tr}\big[A^{\gamma\delta}(  \alpha,\varepsilon)  A^{\delta\gamma}(\beta,\varepsilon)\big]  f_{\gamma}(  \varepsilon)  \left[1-f_{\delta}(  \varepsilon)  \right]\label{S},
\end{eqnarray}
where
\begin{eqnarray}
A^{\gamma\delta}(  \alpha,\varepsilon)  &=&\delta_{\alpha\gamma}\delta_{\alpha\delta}-s^{\alpha\gamma\dag}(  \varepsilon)s^{\alpha\delta}(  \varepsilon)\label{A},
\end{eqnarray}
and $s^{\alpha\beta}(\varepsilon)$ is the scattering matrix for an incident particle of type $\beta$ from the probe to the bound state reflected back as a particle of type $\alpha$. The sign function $\textrm{sgn}(\alpha)$ equals $1$ and $-1$ for $\alpha=e$ and $\alpha=h$, respectively. The Fermi distributions of electron ($e$) and hole ($h$) in the probe are $f_e(\varepsilon)=1/[e^{(\varepsilon-eV)/k_BT}+1]$ and $f_h(\varepsilon)=1/[e^{(\varepsilon+eV)/k_BT}+1]$, respectively, with $T$ being the temperature and $V$ the bias voltage applied to the probe.

Substituting Eq.~\eqref{A} into Eq.~\eqref{I} yields
\begin{eqnarray}
I(V)&=&\frac{e}{h}\int_{0}^{\infty}d\varepsilon\bigg\{  \textrm{Tr}\big[  I_2-T^{ee}(\varepsilon)  +T^{he}(  \varepsilon)  \big]  f_{e}(\varepsilon) -\textrm{Tr}\big[  I_2-T^{hh}(  \varepsilon)+T^{eh}(  \varepsilon)  \big]  f_{h}(  \varepsilon)\bigg\}, \label{I2}
\end{eqnarray}
where $I_2$ denotes a $2\times 2$ identity matrix and $T^{\alpha\beta}(\varepsilon)=s^{\alpha\beta\dag}(\varepsilon)s^{\alpha\beta}(\varepsilon)$. By the sum rule $T^{\alpha\alpha}(  \varepsilon)+T^{\bar\alpha\alpha}(  \varepsilon)=I_2 $, Eq.~\eqref{I2} reduces to
\begin{equation}
I(V)=\frac{2e}{h}\int_{0}^{\infty}d\varepsilon \textrm{Tr}[  T^{he}(\varepsilon)f_{e}(  \varepsilon)-T^{eh}(\varepsilon)f_{h}(  \varepsilon) ]\label{I3}.
\end{equation}
Substituting Eq.~\eqref{A} into Eqs.~\eqref{S} yields
\begin{eqnarray}
S(V)&=&\frac{2e^{2}}{h}\sum_{\alpha,\beta\in e,h}\textrm{sgn}(  \alpha)\textrm{sgn}(  \beta)  \int_{0}^{\infty}d\varepsilon\textrm{Tr}[\delta_{\alpha\beta}I_2-2T^{\beta\alpha}(  \varepsilon)]f_{\alpha}(  \varepsilon)  [  1-f_{\alpha}(\varepsilon)  ], \notag\\
&&+\frac{2e^{2}}{h}\sum_{\alpha,\beta,\gamma,\delta\in e,h}\textrm{sgn}(\alpha)  \textrm{sgn}(  \beta)  \int_{0}^{\infty}d\varepsilon\textrm{Tr}[  s^{\alpha\gamma\dag}(  \varepsilon)  s^{\alpha\delta}(  \varepsilon)  s^{\beta\delta\dag}(  \varepsilon)s^{\beta\gamma}(  \varepsilon) ]f_{\gamma}(\varepsilon) [  1-f_{\delta}(  \varepsilon) ]\label{S3}.
\end{eqnarray}

The scattering matrix of the effective Hamiltonian considered in the main text can be obtained by the Mahaux-Weidenm\"{u}ller formula~\cite{mahaux1969shell}
\begin{eqnarray}
\left(
\begin{array}
[c]{cc}%
s^{ee}\left(  \varepsilon\right)   & s^{eh}\left(  \varepsilon\right)  \\
s^{he}\left(  \varepsilon\right)   & s^{hh}\left(  \varepsilon\right)
\end{array}
\right)  =I_{4}-2\pi iW^{\dag}G^{r}\left(  \varepsilon\right)  W, \label{sm}
\end{eqnarray}
\begin{equation}
G^{r}(  \varepsilon)  =(  \varepsilon I_{2}-H_{B}+i\pi WW^{\dag})  ^{-1}.
\end{equation}
In the basis $\{c^\dag_{k\uparrow},c^\dag_{k\downarrow},c_{k\uparrow},c_{k\downarrow}\}$, one has
\begin{eqnarray}
W=\left(
\begin{array}
[c]{cc}%
W_{e} & W_{h}%
\end{array}
\right),
\end{eqnarray}
\begin{eqnarray}
\hspace{-0.5cm}W_{e}=\sqrt{\rho}\left(
\begin{array}
[c]{cc}%
t_{1\uparrow}^{\ast} & t_{1\downarrow}^{\ast}\\
t_{2\uparrow}^{\ast} & t_{2\downarrow}^{\ast}%
\end{array}
\right),
W_{h}=\sqrt{\rho}\left(
\begin{array}
[c]{cc}%
-t_{1\uparrow} & -t_{1\downarrow}\\
-t_{2\uparrow} & -t_{2\downarrow}%
\end{array}
\right),
\end{eqnarray}
\begin{eqnarray}
H_{B}=\left(
\begin{array}
[c]{cc}%
0 & i\varepsilon_{B}\\
-i\varepsilon_{B} & 0
\end{array}
\right)\label{HB},
\end{eqnarray}
where $\rho$ is the density of states of the probe. Equations \eqref{I3}--\eqref{HB} are used to calculate the transport quantities at finite temperatures.

We proceed to derive the zero-temperature current and current noise, which are presented as Eqs.~(1)--(5) in the main text. At $T=0$, the Fermi distribution function of the electron and hole in the probe become step functions $f_{e}(  \varepsilon)   =\theta(  -\varepsilon+eV)$ and $f_{h}(  \varepsilon)   =\theta(  -\varepsilon-eV)$. As a result, Eqs.~\eqref{I3} and \eqref{S3} reduce to
\begin{eqnarray}
I(V)&=&\frac{2e}{h}\int_{0}^{e|V|}d\varepsilon \textrm{Tr}[  \theta(V)T^{he}(\varepsilon)-\theta(-V)T^{eh}(\varepsilon) ],\label{I4}\\
S(V)&=&\frac{2e^{2}}{h}\sum_{\alpha,\beta,\gamma,\delta\in e,h}\textrm{sgn}(\alpha)  \textrm{sgn}(  \beta)  \int_{0}^{\infty}d\varepsilon\textrm{Tr}\big[  s^{\alpha\gamma\dag}(  \varepsilon)  s^{\alpha\bar\gamma}(  \varepsilon)  s^{\beta\bar\gamma\dag}(  \varepsilon)s^{\beta\gamma}(  \varepsilon)  \big]f_{\gamma}(\varepsilon)  \big[  1-f_{\bar\gamma}(  \varepsilon)  \big],\label{S4}
\end{eqnarray}
where $\bar\gamma$ denotes the index opposite to $\gamma$. Using the orthogonality relation $s^{\alpha{\gamma}}(  \varepsilon)  s^{\beta{\gamma}\dag}(\varepsilon)+s^{\alpha\bar{\gamma}}(  \varepsilon)  s^{\beta\bar{\gamma}\dag}(
\varepsilon)=\delta_{\alpha\beta}I_2$ and the sum rule $T^{\alpha\beta}(\varepsilon)+T^{\bar\alpha\beta}(\varepsilon)=I_2$, Eq.~\eqref{S4} is recast as
\begin{eqnarray}
S(V)&=&\frac{8e^{2}}{h}\sum_{\gamma\in e,h}\int_{0}^{\infty}d\varepsilon\textrm{Tr}[  T^{e\gamma}(  \varepsilon)  T^{h\gamma}(  \varepsilon)  ]
f_{\gamma}(  \varepsilon)[  1-f_{\bar{\gamma}}(  \varepsilon) ].\label{S5}
\end{eqnarray}
The particle-hole symmetry of the Hamiltonian $H=-CHC^{-1}$, with $C=\tau_x K$, leads to $s^{\alpha\beta}(\varepsilon)=[s^{\bar\alpha\bar\beta}(-\varepsilon)]^*$ and
$T^{\alpha\beta}(\varepsilon)=[T^{\bar\alpha\bar\beta}(-\varepsilon)]^T$, by which Eqs.~\eqref{I4} and \eqref{S5} can be simplified to
\begin{eqnarray}
I(V)&=&\frac{2e}{h}\int_{0}^{eV}d\varepsilon \textrm{Tr}[  T^{he}(\varepsilon) ],\label{I5}\\
S(V)&=&\textrm{sgn}(V)\frac{8e^{2}}{h}\int_{0}^{eV}d\varepsilon\textrm{Tr}\big\{T^{he}(  \varepsilon)[I_2-T^{he}(\varepsilon)] \big\}\label{S6},
\end{eqnarray}
where $T^{ee}(\varepsilon)$ has been replaced with $I_2-T^{he}(\varepsilon)$ due to the sum rule.

It follows from Eq.~\eqref{sm} that $s^{he}(  \varepsilon)  =-2\pi iW_{h}^{\dag}G^{r}(\varepsilon)  W_{e}$, therefore,
\begin{eqnarray}
\textrm{Tr}[  T^{he}(  \varepsilon) ]&=&\textrm{Tr}[  s^{he\dag}(  \varepsilon)  s^{he}(\varepsilon) ]=\textrm{Tr}[\widetilde{T}(\varepsilon) ],\label{trace1}\\
\widetilde{T}(\varepsilon)&=&\Gamma_{e}G^{a}(  \varepsilon)  \Gamma_{h}G^{r}(\varepsilon),\label{tildeT}
\end{eqnarray}
where $\Gamma_{\alpha}=2\pi W_{\alpha}W_{\alpha}^\dag$ and $G^a(\varepsilon)=G^{r\dag}(\varepsilon)$. Similarly,
\begin{equation}
\textrm{Tr}\{ T^{he}(  \varepsilon)[I_2-T^{he}(  \varepsilon)]\} =\textrm{Tr}\{\widetilde{T}(\varepsilon)[I_2-\widetilde{T}(\varepsilon)] \}\label{trace2}.
\end{equation}
In terms of the eigen values $T_\pm(\varepsilon)$ of the $2\times 2$ matrix $\widetilde{T}(\varepsilon)$, Eqs.~\eqref{trace1} and \eqref{trace2} can be rewritten as
\begin{eqnarray}
\hspace{-0.5cm}\textrm{Tr}[  T^{he}(  \varepsilon) ]&=&\sum_{s=\pm}T_s(\varepsilon) ,\label{trace3}\\
\hspace{-0.5cm}\textrm{Tr}\{ T^{he}(  \varepsilon)[I_2-T^{he}(  \varepsilon)]\}&=&\sum_{s=\pm}T_s(\varepsilon)[1-T_s(\varepsilon)]\label{trace4}.
\end{eqnarray}
Substituting Eqs.~\eqref{trace3} and \eqref{trace4} into Eqs.~\eqref{I5} and \eqref{S6} yields Eqs.~(1) and (2) in the main text.

With the parametrization scheme $t_{1\uparrow} =t_{1}\sin(\theta_1/2)$, $t_{1\downarrow}=-t_{1}\cos(\theta_1/2)$, $t_{2\uparrow}=-it_{2}\sin(\theta_2/2)$, $t_{2\downarrow}=-it_{2}\cos(\theta_2/2)$ in the main text, one has
\begin{eqnarray}
{\Gamma}_{e}=\Gamma\left(
\begin{array}
[c]{cc}%
1-r & i\sqrt{r(  1-r)  }\cos\theta\\
-i\sqrt{r(  1-r)  }\cos\theta & r
\end{array}
\right),\label{ge}\\
{\Gamma}_{h}=\Gamma\left(
\begin{array}
[c]{cc}%
1-r & -i\sqrt{r(  1-r)  }\cos\theta\\
i\sqrt{r(  1-r)  }\cos\theta & r
\end{array}
\right),\label{gh}\\
G^{r}(  \varepsilon)  =\left(
\begin{array}
[c]{cc}%
\varepsilon+i\Gamma(  1-r)   & -i\varepsilon_{B}\\
i\varepsilon_{B} & \varepsilon+i\Gamma r
\end{array}
\right)  ^{-1}\label{gr},
\end{eqnarray}
where $\Gamma=\Gamma_{1}+\Gamma_{2}$, $r=\Gamma_{2}/\Gamma$,  $\Gamma_{1,2}=2\pi\rho t_{1,2}^{2}$, and $\theta=(  \theta_{1}+\theta_{2})  /2$.
Substituting Eqs.~\eqref{ge}, \eqref{gh}, and \eqref{gr} into Eq.~\eqref{tildeT} yields $\widetilde{T}(\varepsilon)$, whose eigen values $T_{\pm}(\varepsilon)$ can be analytically solved and are presented as Eqs.~(3)--(5) in the main text.

As mentioned in the main text, a zero-bias conductance peak (ZBCP) in a differential conductance curve $G(V)$ may also arise from a probe simultaneously coupled to $N>1$ near-zero-energy superconducting bound states. This situation can be modeled by extending the Hamiltonian $H_B$ and $H_T$ in the main text to
\begin{eqnarray}
H_{B}&=&i\sum_{n=1}^{N}\varepsilon_{Bn} \gamma_{n1}\gamma_{n2},\label{S25}\\
H_{T}&=&\sum_{n=1}^N\sum_{m=1}^{2}\sum_{ k\sigma}(  t_{nm\sigma}c_{k\sigma}^{\dag}-t_{nm\sigma}^{\ast}c_{k\sigma})  \gamma_{nm},\label{S26}
\end{eqnarray}
where $n$ indexes different bound states. The associated $W_{e(h)}$, $W^\dag_{e(h)}$, $G^{r(a)}(\varepsilon)$ are $2N \times 2$, $2 \times 2N$, and $2N \times 2N$ matrices, respectively. It follows from Eq.~\eqref{trace1} that $T^{he}(\varepsilon)$ is a $2\times 2$ matrix, indicating that at most two eigen channels are available for transport. 

\section{Fano factors in the large-bias and small-bias limits}\label{sec2}
\begin{figure*}[htbp]
\centering
\includegraphics[width=0.6\columnwidth]{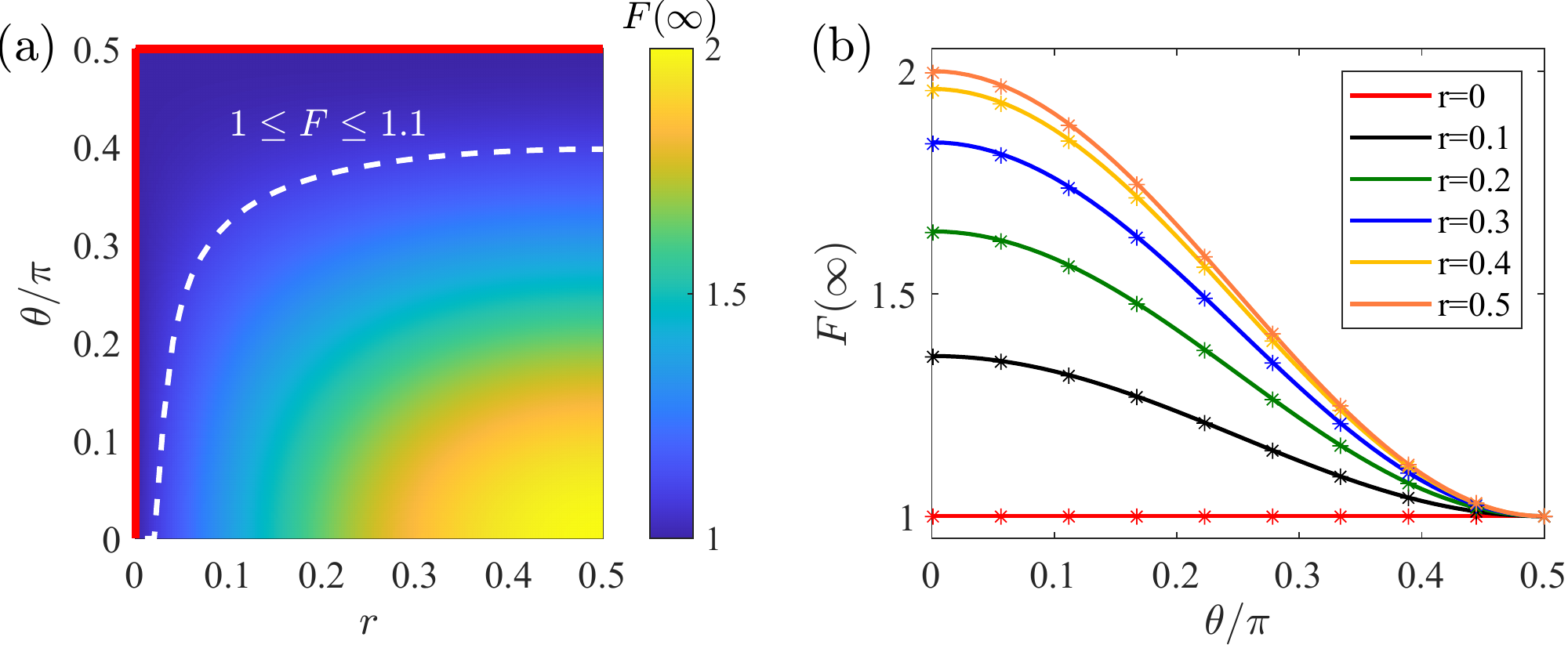}
\caption{(a) Zero-temperature Fano factor $F$ in the large-bias limit ($V\rightarrow\infty$) as functions of $r$ and $\theta$. $F=1$ along the two red solid lines and $1\le F\le 1.1$ in the region above the white dashed line. (b) Thermal effects on $F(\infty)$. Solid lines are linecuts of (a) at different $r$ values, while asterisks show the corresponding results calculated at a finite temperature $k_B T=0.2~\Gamma$. }\label{FigS0_largeV}
\end{figure*}

The Fano factor of MBS/QMBS is predicted~\cite{perrin2022identifying} to be $1$ when $e|V|$ (where $V$ is the bias voltage) is much larger than tunneling broadening and lower than the superconducting gap. This inspires a recent STM experiment~\cite{ge2023single} to distinguish MBS/QMBS from ABSs by comparing the Fano factor in conventional superconductor NbSe$_2$ and iron-based superconductor FeTe$_{0.55}$Se$_{0.45}$. Specifically, ZBCPs are observed in the vortices of both materials, however, the corresponding Fano factors $F(\infty)$ measured at a large bias voltage are both close to 1. This unexpected result is not surprising in view of Fig.~\ref{FigS0_largeV}(a) plotted after $F(\infty)=1$ for $r=0$ and $F(\infty)=1+4r(1-r)\cos^2\theta$ for $r>0$, which are predicted by our Eqs.~(1)--(5) in the main text along with the definition $F(\infty)=\textrm{lim}_{V\rightarrow\infty} S(V)/2e|I(V)|$. As we can see, zero-energy ABSs can also lead to a large-bias Fano factor equal or close to $1$: $F(\infty)=1$ along the red solid lines and $1\le F(\infty)\le 1.1$ in the region above the white dashed line. Therefore, the ZBCPs observed in both conventional and iron-based superconductors~\cite{ge2023single} may be caused by zero-energy ABSs.

In practice, transport experiments are carried out at finite temperatures. The nearly quantized zero-bias conductance peaks (ZBCPs) in local tunneling spectroscopy are observed at $T=50$ mK and $T=377$ mK, respectively, in experiments of hybrid InAs-Al nanowire~\cite{wang2022plateau} and iron-based superconductor FeTe$_{0.55}$Se$_{0.45}$~\cite{zhu2020nearly}. The probe-induced broadenings of the ZBCPs in both experiments are estimated to be $\Gamma=0.15$ meV, corresponding to the ratio $k_B T/\Gamma=0.03$ and $0.27$, respectively.

\begin{figure*}[htbp]
\centering
\includegraphics[width=\columnwidth]{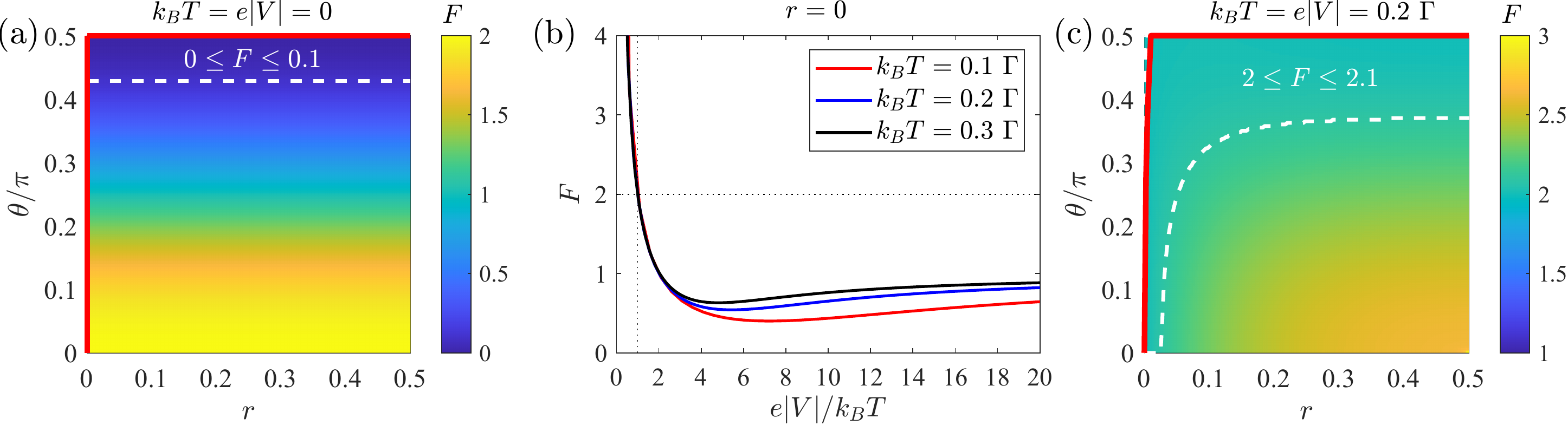}
\caption{(a) Zero-temperature Fano factor $F$ in the zero-bias limit as functions of $r$ and $\theta$. $F=0$ along the two red solid lines and $0\le F\le 0.1$ above the white dashed line. (b) Fano factor as a function of $e|V|/k_B T$ at different nonzero temperatures. Remarkably, $F=2$ at $e|V|/k_BT=1$. (c) Similar to (a) but for $k_BT=e|V|=0.2~\Gamma$. $F=2$ along the two red solid lines and $2\le F\le 2.1$ between the red solid and the white dashed lines.}\label{FigS0_smallV}
\end{figure*}

To inspect the thermal effect on the large-bias Fano factor, we show in Fig.~\ref{FigS0_largeV}(b) the linecuts of Fig.~\ref{FigS0_largeV}(a) at different $r$ values (solid lines) and  the corresponding results calculated at $k_BT/\Gamma=0.2$ (asterisks). As we can see, the thermal fluctuation has no influence on the large-bias Fano factor. This can be attributed to the fact that at a large enough bias $e|V|\gg k_B T$ both current ($I$) and current noise ($S$) are saturated independent of the temperature and so is the corresponding Fano factor defined as $F=S/2e|I|$.

Below, we shall demonstrate that detecting the small-bias Fano factor can not distinguish MBS/QMBS from trivial zero-energy ABSs as well. At zero temperature, a theory~\cite{golub2011shot} predicted that the Fano factor of an MBS/QMBS is $0$ in the zero-bias limit.
However, it follows from Eqs.~(1)--(5) in the main text that the Fano factor in the zero-bias limit is $F=2\cos^2\theta$, as plotted in Fig.~\ref{FigS0_smallV}(a). This invalidates taking a vanishing zero-bias Fano factor as a signature of MBS/QMBS, since $F=0$ along two red solid lines and $0\le F\le 0.1$ above the white dashed line. At finite temperatures, as shown in Fig.~\ref{FigS0_smallV}(b), the Fano factor for $e|V|<k_BT$ diverges since the voltage-driven current is small compared to the thermal-dominated current noise. Interestingly, the Fano factor of MBS/QMBS ($r=0$) becomes integer $2$ at $e|V|/k_BT=1$, in contrast to $0$ predicted at zero temperature, as we will discuss below. Figure \ref{FigS0_smallV}(b) also shows that as $|V|$ increases to the large-bias limit, the Fano factor converges to $1$, which agrees with Fig.~\ref{FigS0_largeV}(a). In Fig.~\ref{FigS0_smallV}(c), we show the Fano factor at $e|V|=k_B T=0.2 \Gamma$. As we can see, $F=2$ along two red solid lines and $2\le F\le 2.1$ in the area within the white dashed line, indicating that detecting the Fano factor at small bias also can not distinguish MBS/QMBS from zero-energy ABSs.

As mentioned above, the zero-bias Fano factor of MBS/QMBS at zero temperature is $0$, in contrast to the integer $2$ at finite temperatures. The latter corresponds to the effective charge of $2e$ in the Andreev reflection process as expected. This seeming discrepancy can be explained as follows. In Ref.~\cite{golub2011shot} and in our main text, the Fano factor is defined as $F(V)\equiv S(V)/2e|I(V)|$. However, the Fano factor defined in this way conveys only the quasi-particle information in the weak tunneling limit.
Indeed, near the perfect tunneling, the current approaches its perfect expression $I(V) \approx V2e^2/h$, as a linear function in $V$.
By contrast, the noise is produced by the fluctuation of the current, and thus closely related to the deviation from the perfect tunneling current (i.e., the backscattering current defined as $I(V) - V2e^2/h$).
Consequently, to obtain the quasi-particle information, the reasonable way is to define the Fano factor as the ratio between the total noise, and the source of noise, i.e., the backscattering current that represents the current fluctuation.
Indeed, for a non-interacting system, the knowledge of Fermi-liquid tells us that both the backscattering current and noise are both cubic in the bias voltage. Only the ratio between them is meaningful. We thus define another Fano factor as
\begin{equation}
    \delta F(V)\equiv \frac{S(V)}{2e \big|I(V) - \frac{2e^2}{h}V \big|}.
    \label{eq:delta_f}
\end{equation}
This definition of Fano factor has been used actually in e.g., Ref.~\cite{sela2006fractional} to extract the effective charge of the quasi-particle of a Kondo model near the perfect tunneling regime.

For $\varepsilon_{B}=0$ and $T=0$, analytical expressions of current and current noise are available
\begin{eqnarray}
I(V)&=&\frac{2e\Gamma}{h}\bigg\{r[1-2(1-r)\cos^2\theta]\tan^{-1}\frac{eV}{r\Gamma}+(1-r)(1-2r\cos^{2}\theta)\tan^{-1}\frac{eV}{(1-r)\Gamma}\bigg\},\label{iiv}\\
S(V)&=&\textrm{sgn}(  V)  \frac{4e^{2}\Gamma}{h}\bigg\{-\Gamma eV\frac{(  1-r)  ^{2}(  1-2r\cos^{2}\theta)^{2}}{(  eV)  ^{2}+(  1-r)  ^{2}\Gamma^{2}}-\Gamma eV\frac{r^{2}[  1-2(  1-r)  \cos^{2}\theta]^{2}}{(  eV)  ^{2}+r^{2}\Gamma^{2}}\notag\\
&&+(  1-r) [  1-8(  1-r)  r^{2}\cos^{4}\theta]  \tan^{-1}\frac{eV}{(  1-r)  \Gamma}+r[  1-8r(  1-r)  ^{2}\cos^{4}\theta]  \tan^{-1}\frac{eV}{r\Gamma}\bigg\}\label{ssv}.
\end{eqnarray}
It follows from Eq.~\eqref{ssv} that
\begin{equation}
P(V)\equiv \bigg\vert \frac{dS(V)}{dV}\bigg\vert=\frac{C_1}{{(  eV)  ^{2}+r  ^{2}\Gamma^{2}}}+\frac{C_2}{{(  eV)  ^{2}+(  1-r)  ^{2}\Gamma^{2}}}+\frac{C_3}{{[(  eV)  ^{2}+r  ^{2}\Gamma^{2}]^2}}+\frac{C_4}{{](  eV)  ^{2}+(  1-r)  ^{2}\Gamma^{2}]^2}},\label{PV_anal}
\end{equation}
where $C_i$ ($i=1,2,3,4$) are coefficients. In the small-$r$ limit, Eq.~\eqref{PV_anal} implies a Lorentzian peak with a width of $r\Gamma$ in the $P(V)$ curves, as shown in Fig.~2(b) in the main text.

For the MBS/QMBS case ($r=0$), Eqs.~\eqref{iiv} and \eqref{ssv} reduce to
\begin{eqnarray}
I(V) &=& \frac{2e\Gamma}{h} \tan^{-1}\frac{eV}{\Gamma},\label{eq:zero_t_current}\\
S(V)&=&\textrm{sgn}(  V)  \frac{4e^{2}\Gamma}{h}\bigg[  \tan^{-1}\frac{eV}{\Gamma}-\frac{eV\Gamma}{(  eV)  ^{2}+\Gamma^{2}}\bigg],
\label{eq:zero_t_noise}
\end{eqnarray}
which agree with those in Ref.~\cite{golub2011shot}.
For a small bias voltage, one can expand Eqs.~\eqref{eq:zero_t_current} and \eqref{eq:zero_t_noise} to the leading order of $V$ as
\begin{equation}
     I(V) \approx  \frac{2e}{h} \bigg[ eV - \frac{(eV)^3}{3\Gamma^2} \bigg], \\  ~~ S(V) \approx \frac{8}{3} \frac{e^2}{h} \frac{(eV)^3}{\Gamma^2}.
\end{equation}
As we mentioned above, both the current noise and the backscattering part of the current are cubic in $V$, indicating that they are introduced by the same source: the fluctuation of the tunneling current when approaching the perfect transmitting point.
In addition, following its definition, Eq.~\eqref{eq:delta_f}, the Fano factor defined with the backscattering current equals 2 in the near-perfect tunneling regime. This result indicates that two electrons are transferred simultaneously when MBS/QMBS induced resonant Andreev reflection occurs.

\section{$G(V)$ and $P(V)$ correlations for zero-energy ABSs with unquantized ZBCPs}\label{sec3}
\begin{figure*}[htbp]
\centering
\includegraphics[width=0.9\columnwidth]{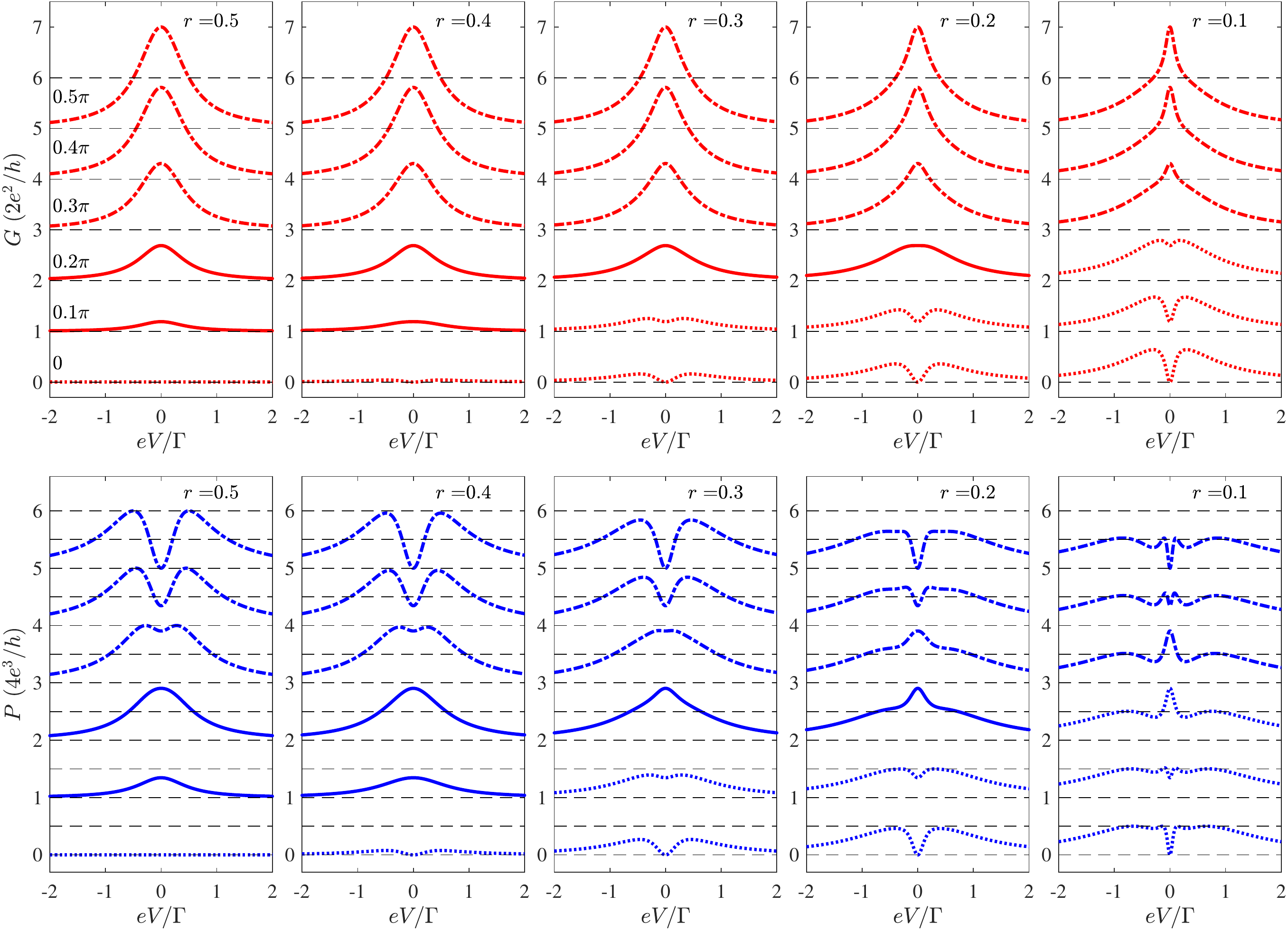}
\caption{Zero-temperature $G(V)$ and $P(V)$ curves of zero-energy ABSs with different $(r,\theta)$ combinations. The lineshapes of $G(V)$ curves can be classified into three kinds: a zero-bias peak lager (dash-dotted lines) or smaller (solid lines) than $2e^2/h$, and a zero-bias dip (dotted lines). The $P(V)$ curves are plotted with the same line styles as the corresponding $G(V)$ curves. The curves for different $\theta$ values are offset by one unit for clarity.}\label{Fig_scan}
\end{figure*}

In Fig.~2 in the main text, we have studied the $G(V)$ and $P(V)$ curves of zero-energy ABSs with $\theta=0.25\pi$ and different $r$ values. In Fig.~\ref{Fig_scan}, we show the $G(V)$ and $P(V)$ curves of zero-energy ABSs with different $(r,\theta)$ combinations. In the top panel, each column is calculated with a fixed $r$ value and different $\theta$ values as indicated in the first column. As we can see, the lineshapes of $G(V)$ curves can be classified into three kinds: a zero-bias peak lager (dash-dotted lines) or smaller (solid lines) than $2e^2/h$, and a zero-bias dip (dotted lines). Note that $G(0)$ is independent of the $r$ value, in agreement with the analytical expression $G(0)=4e^2\sin^2\theta/h$ predicted by Eqs.~(1)--(5) in the main text. Physically, the zero-bias peak (dip) results from the constructive (destructive) interference of the charge tunnelings between the probe and the two Majorana components $\gamma_{1,2}$ of an ABS. Specifically, for $\theta=0$, $t_{1\uparrow}=t_{2\uparrow}=0$ (or $t_{1\downarrow}=t_{2\downarrow}=0$), $\gamma_1$ and $\gamma_2$ couple to a common spin channel of the probe, their competition leads to $G(0)=0$; for $\theta=0.5\pi$, $t_{1\downarrow}=t_{2\uparrow}=0$ (or $t_{1\uparrow}=t_{2\downarrow}=0$), $\gamma_1$ and $\gamma_2$ couple to opposite spin channels, resonant Andreev reflection occurring between each Majorana component and its coupled spin channel are additive, thus leading to a total conductance $G(0)=4e^2/h$. In the bottom panel, we present the corresponding $P(V)$ curves with identical line styles of $G(V)$ curves. Overall, in agreement with Figs.~4(e) and 4(f) in the main text, $P_{max}> 2e^3/h$ when the maximum of a $G(V)$ curve exceeds $0.29 \times 2e^2/h$. Particularly, $P_{max}> 2e^3/h$ is most remarkable for the nearly quantized ZBCPs.

\section{Supplemental information of Fig.~3 and Fig.~4 in the main text}\label{sec4}
\begin{figure}[htbp]
\centering
\includegraphics[width=\columnwidth]{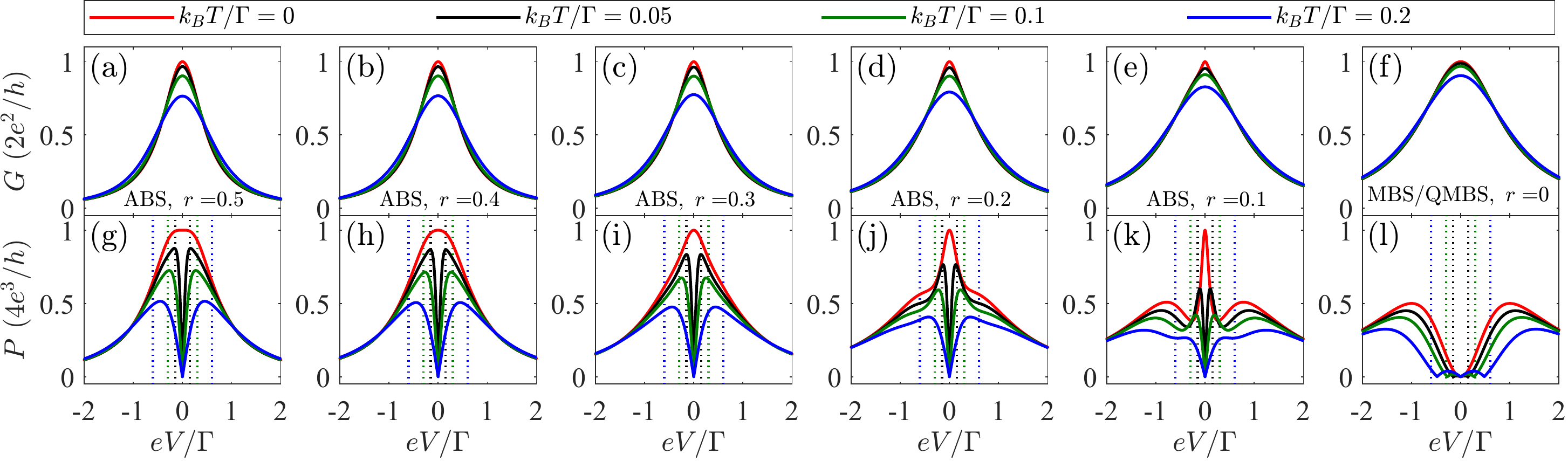}
\caption{Temperature impact on the $G(V)$ and $P(V)$ curves in Figs.~2(a) and 2(b) in the main text. }\label{Fig_variousToneboundstate}
\end{figure}

In Figs.~3(a)--3(d) in the main text, we present the temperature impact on the $G(V)$ and $P(V)$ curves of zero-energy ABSs with $r=0$ and $0.4$. For completeness, we show the results of other $r$ values in Fig.~\ref{Fig_variousToneboundstate}. 

\begin{figure*}[htbp]
\centering
\includegraphics[width=0.4\columnwidth]{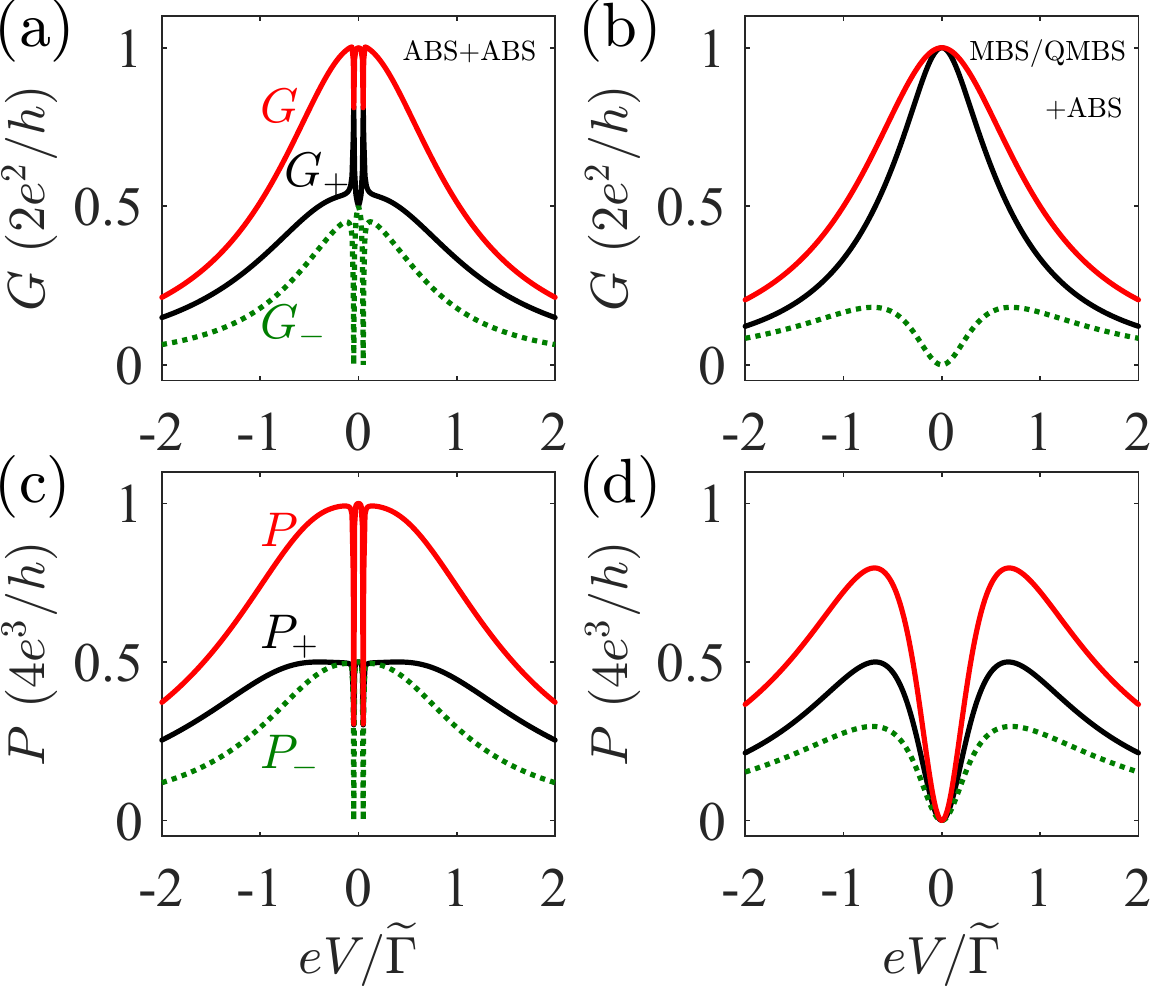}
\caption{Contributions from the two eigen channels ``$\pm$'' to the zero-temperature $G(V)$ and $P(V)$ curves shown in Figs.~4(a)--4(d) in the main text.}\label{Fig_zeroTtwoboundstate}
\end{figure*}

As indicated by Eqs.~(1) and (2) in the main text, $G(V)$ and $P(V)$ are contributed independently from two eigen channels ``$\pm$''. We show in Fig.~\ref{Fig_zeroTtwoboundstate} how the two channels contribute to the zero-temperature $G(V)$ and $P(V)$ curves in Figs.~4(a)--4(d) in the main text, for the cases where a probe is simultaneously coupled to a pair of low-energy ABSs or to a combination of an MBS/QMBS and zero-energy ABS. It is evident that both channels are involved in the transport,  $G_{max}>0.29\times 2e^2/h$, and $P_{max}>2e^3/h$. These are in agreement with the predictions of Figs.~4(e) and 4(f) in the main text.

\begin{table}[htbp]
\centering
\tabcolsep=0.6cm
\caption{Figures 4(a)--4(d) in the main text are calculated by Hamiltonian \eqref{S25} and \eqref{S26} with the listed tunneling amplitudes. $\alpha$ and $\beta$ denote the two low-energy bound states coupled to the probe. These amplitudes are in units of $t_{\alpha1\uparrow}$.}\label{tb1}
\begin{tabular}{ccccccccc}
\hline
\hline
&$t_{\alpha1\uparrow}$ &$t_{\alpha2\uparrow}$ &$t_{\alpha1\downarrow}$ &$t_{\alpha2\downarrow}$ &$t_{\beta1\uparrow}$ &$t_{\beta2\uparrow}$ &$t_{\beta1\downarrow}$ &$t_{\beta2\downarrow}$\\
\hline
Figs.~4(a) and 4(c) &$1$ &$0.3i$ &$0.4$ &$0.7i$ &$0.5$ &$0.5i$ &$0.15$ &$1.1i$ \\
\hline
Figs.~4(b) and 4(d) &$1$ &$0$ &$0$ &$0$ &$0.3$ &$-i$ &$0.7$ &$0.7i$ \\
\hline
\hline
\end{tabular}
\end{table}

In Table \ref{tb1}, we list the tunneling amplitudes, between the probe and the two Majorana components (denoted as $1$ and $2$, respectively) of the two bound states (denoted as $\alpha$ and $\beta$, respectively), used for the calculations of Figs.~4(a)--4(d) in the main text. As shown in Fig.~1 in the main text, the probe is coupled to both Majorana components of an ABS while to only one Majorana component of an MBS/QMBS. For this reason, we set all tunneling amplitudes to non-zeros for calculating Figs.~4(a) and 4(c), and three tunneling amplitudes associated with bound state $\alpha$ to zeros for calculating Figs.~4(b) and 4(d). It is noteworthy to point out an accidental case in which two isolated zero-energy ABSs coupled to a common probe can combine into an MBS/QMBS and an ABS. To be specific, if $t_{\alpha2\uparrow}=t_{\beta2\uparrow}$ and $t_{\alpha2\downarrow}=t_{\beta2\downarrow}$, the tunneling Hamiltonian  Eq.~\eqref{S26} can be explicitly written as
\begin{equation}
H_{T}=\sum_{k}\left[t_{\alpha1\uparrow}c_{k\uparrow}^{\dag}\gamma_{\alpha1}+t_{\beta1\uparrow
}c_{k\uparrow}^{\dag}\gamma_{\beta1}+t_{\alpha1\downarrow}c_{k\downarrow
}^{\dag}\gamma_{\alpha1}+t_{\beta1\downarrow}c_{k\downarrow}^{\dag}%
\gamma_{\beta1}+t_{\alpha2\uparrow}c_{k\uparrow}^{\dag}\left(  \gamma_{\alpha2}+\gamma
_{\beta2}\right)  +t_{\alpha2\downarrow}c_{k\downarrow}^{\dag}\left(
\gamma_{\alpha2}+\gamma_{\beta2}\right)  +H.c.\right],
\end{equation}
Under a unitary transformation, $\gamma_{\alpha2}+\gamma_{\beta2}$ will combine into a new Majorana component coupled to the probe while the other Majorana component $\gamma_{\alpha2}-\gamma_{\beta2}$ is decoupled. Therefore, in this accidental case, the $P(V)$ curve will exhibit a zero-bias dip and $P_{max}>2e^3/h$, just like Fig.~\ref{Fig_zeroTtwoboundstate}(d).

\section{Details of the numerical simulations of hybrid nanowires}\label{sec5}
Following Refs.~\cite{liu2017andreev,pan2020physical}, we perform numerical simulations of semiconductor-superconductor hybrid nanowires using the Hamiltonian  
\begin{equation}
H=\Big[-\frac{\hbar^2}{2m^\ast}\partial_x^2-i\alpha\partial_x\sigma_y-\mu+V_\textrm{dis}(x)\Big]\tau_z+V_Z\sigma_x+\Sigma(\varepsilon)+i\eta,
\end{equation}
with 
\begin{equation}
\Sigma(\varepsilon)=-\lambda\frac{\varepsilon+\Delta(V_Z)\tau_x}{\sqrt{\Delta^2(V_Z)-\varepsilon^2}},\\ ~~~~\Delta(V_Z)=\Delta_0\sqrt{1-(V_Z/V_C)^2}\theta(V_C-V_Z),
\end{equation}
where $m^\ast$, $\alpha$, $\mu$, and $V_Z$ are the semiconductor's effective electron mass, spin-orbit coupling strength, chemical potential, and Zeeman energy induced by a magnetic field along the nanowire, respectively. This Hamiltonian is a refined version of the minimal 1D model~\cite{lutchyn2010majorana,oreg2010helical} that has been extensively employed to study hybrid nanowire systems. It explicitly takes into account several physical effects in realistic experiments: (i) $\Sigma(\varepsilon)$ accounts for the superconducting proximity effect on the semiconductor,  with $\lambda$ being the coupling strength between the semiconductor and superconductor; (ii) $\Delta(V_Z)$ accounts for the collapse of superconductor's bulk gap induced by a magnetic field, with $\Delta_0$ being the superconductor paring potential at $V_Z=0$ and $V_C$ being the critical Zeeman energy above which the superconductor's bulk gap completely collapses; (iii) $V_\textrm{dis}(x)$ describes a short-range disorder in the chemical potential along the nanowire, which is generated by an array of uncorrelated Gaussian random numbers with a mean value of $0$ and a standard deviation of $\sigma$; and (iv) $i\eta$ represents a dissipation due to the coupling between the nanowire and its surrounding environment~\cite{liu2017role}. Lastly, as pointed out in Ref.~\cite{liu2017andreev}, multi-subband nanowires can be effectively modeled by a few isolated single-subband nanowires with different chemical potentials.

\begin{table}[htbp]
\centering
\tabcolsep=0.25cm
\caption{Parameters used for the simulations shown in Fig.~5 in the main text and in Sec.~\ref{sec5}. $L$ is the length of the hybrid nanowire, $\alpha$ is the spin-orbit coupling strength, $\mu$ is the chemical potential, $V_\textrm{dis}(x)$ is the disorder in chemical potential along the hybrid nanowire, and $\eta$ is the dissipation strength. }\label{tb2}
\begin{tabular}{cccccc}
\hline
\hline
&$L~\textrm{(nm)}$ &$\alpha~\textrm{(meVnm})$ &$\mu~\textrm{(meV)}$ &$V_\textrm{dis}(x)$ &$\eta~\textrm{(meV)}$ \\
\hline
wire $\#1$, Figs.~5(a) and 5(c) &3000 &50 &0.4 &0 &0.0001 \\
\hline
wire $\#2$, Figs.~5(b) and 5(d) &1000 &30 &0.993 &Fig.~\ref{Fig_randnumber}(a) &0.0001 \\
\hline
wire $\#1$, Figs.~\ref{Fig_MBS_dissipation}(a) and \ref{Fig_MBS_dissipation}(c) &3000 &50 &0.4 &0 &0.0005 \\
\hline
wire $\#1$, Figs.~\ref{Fig_MBS_dissipation}(b) and \ref{Fig_MBS_dissipation}(d) &3000 &50 &0.4 &0 &0.005 \\
\hline
wire $\#3$, Figs.~\ref{Fig_ABS_dissipation}(a) and \ref{Fig_ABS_dissipation}(c) &1000 &30 &0.978 &Fig.~\ref{Fig_randnumber}(b) &0.0001 \\
\hline
wire $\#3$, Figs.~\ref{Fig_ABS_dissipation}(b) and \ref{Fig_ABS_dissipation}(d) &1000 &30 &0.978 &Fig.~\ref{Fig_randnumber}(b) &0.01 \\
\hline
wire $\#4$, Figs.~\ref{Fig_ABS_and_background}(a) and \ref{Fig_ABS_and_background}(b) &1000 &30 &\makecell[c]{ 1st subband: 1 \\ 2nd subband: 3} &\makecell[c]{ 1st subband: Fig.~\ref{Fig_randnumber}(c)\\ 2nd subband: 0} &0.0001 \\
\hline
wire $\#5$, Figs.~\ref{Fig_ABS_and_ABS}(a) and \ref{Fig_ABS_and_ABS}(b) &1000 &30 &\makecell[c]{1st subband: 1 \\ 2nd subband: 1.79} &\makecell[c]{1st subband: Fig.~\ref{Fig_randnumber}(c) \\ 2nd subband: Fig.~\ref{Fig_randnumber}(c)} &0.0001 \\
\hline
\hline
\end{tabular}
\end{table}

Our simulations are conducted on a one-dimensional lattice with the lattice constant of $10$ nm. To simulate the tunneling conductance and differential current noise, we add a $\delta$-type tunnel barrier with a height of $10$ meV to the leftmost lattice site and attach the left end of the hybrid nanowire to a probe described by $H_P$ in the main text. All simulations use the reasonable parameters as $m^\ast=0.015~m_e$, $\Delta_0=0.2$ meV, $\lambda=0.2$ meV, $\sigma=2$ meV, and $V_C=0.85$ meV, while the other parameters are listed in Table \ref{tb2}.

\begin{figure*}[htbp]
\centering
\includegraphics[width=0.8\columnwidth]{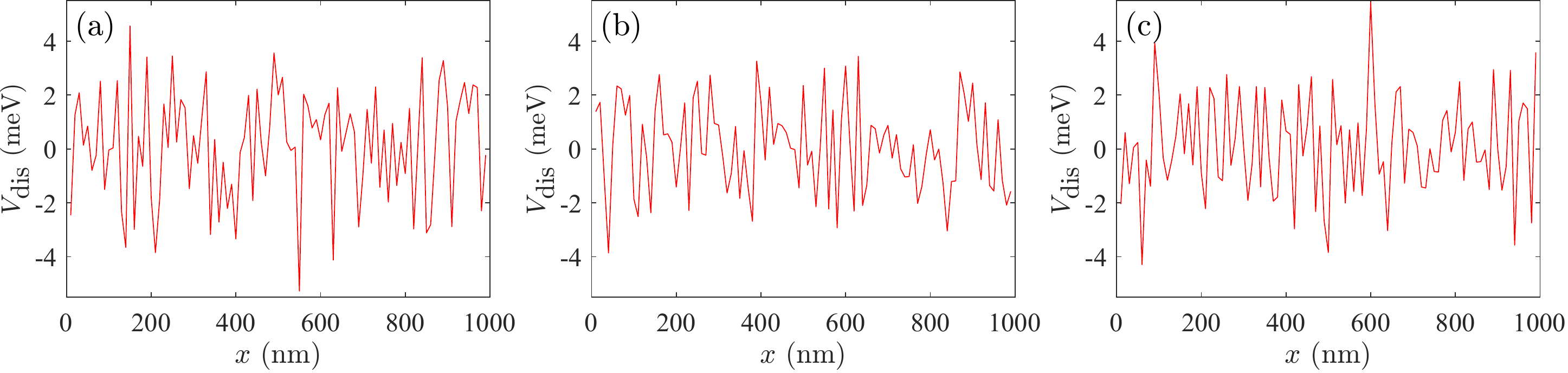}
\caption{Three configurations of disorder in chemical potential along a hybrid nanowire with a length of $1000$ nm. They are used in the simulations summarized in Table \ref{tb2}. We note that not all disorder configurations can result in a nearly quantized conductance plateau. Actually, each configuration is postselected from about one hundred simulations. Such a postselection procedure has been employed in Ref.~\cite{sarma2021disorder}.}\label{Fig_randnumber}
\end{figure*}

The local density of states of the hybrid nanowire can be calculated by
\begin{eqnarray}
\rho(x,\varepsilon)&=&-\frac{1}{\pi}\textrm{Im}\big[G^r_{11}(x,\varepsilon)+G^r_{22}(x,\varepsilon)\big],\\
G^r(x,\varepsilon)&=&\langle x\vert (\varepsilon-H)^{-1}\vert x\rangle.
\end{eqnarray}
We assume that the bias voltage $V$ is applied to the probe while the hybrid nanowire is grounded. At zero temperature, the tunneling conductance and differential current noise can be calculated by~\cite{anantram1996current}
\begin{eqnarray}
G(V)&=&\frac{e^2}{h}  \textrm{Tr}\big[I_2-T_{NN}^{ee}(\varepsilon)+T^{he}_{NN}(\varepsilon)\big],\\
P(V)&=&\frac{2e^3}{h} \textrm{Tr}\big\{[I_2-T_{NN}^{ee}(\varepsilon)]T_{NN}^{ee}(\varepsilon)+[I_2-T^{he}_{NN}(\varepsilon)]T^{he}_{NN}(\varepsilon)+2T^{he}_{NN}(\varepsilon)T^{ee}_{NN}(\varepsilon)\big\},
\end{eqnarray}
where $T_{NN}^{\alpha\beta}(\varepsilon)=s_{NN}^{\alpha\beta\dag}(\varepsilon)s_{NN}^{\alpha\beta}(\varepsilon)$ and the scattering matrix $s_{NN}^{\alpha\beta}(\varepsilon)$ can be obtained by the Kwant package~\cite{groth2014kwant}. 

\begin{figure*}[htbp]
\centering
\includegraphics[width=0.6\columnwidth]{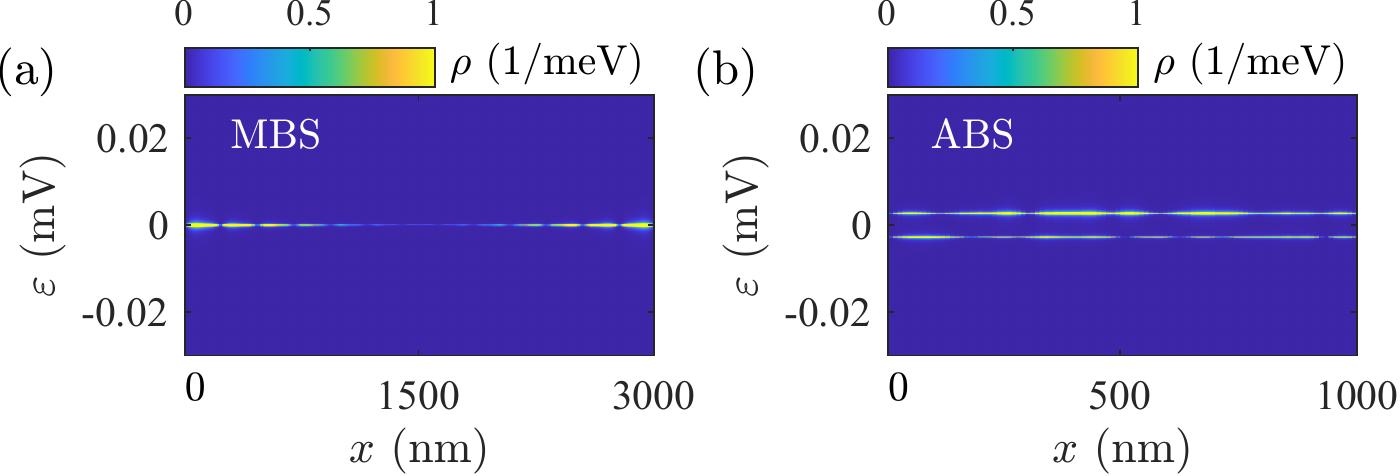}
\caption{Local density of states $\rho(x,\varepsilon)$ of the bound states indicated by the colored bars in Figs.~5(a) and 5(b), respectively, in the main text.}\label{Fig_DOS}
\end{figure*}

\begin{figure*}[htbp]
\centering
\includegraphics[width=\columnwidth]{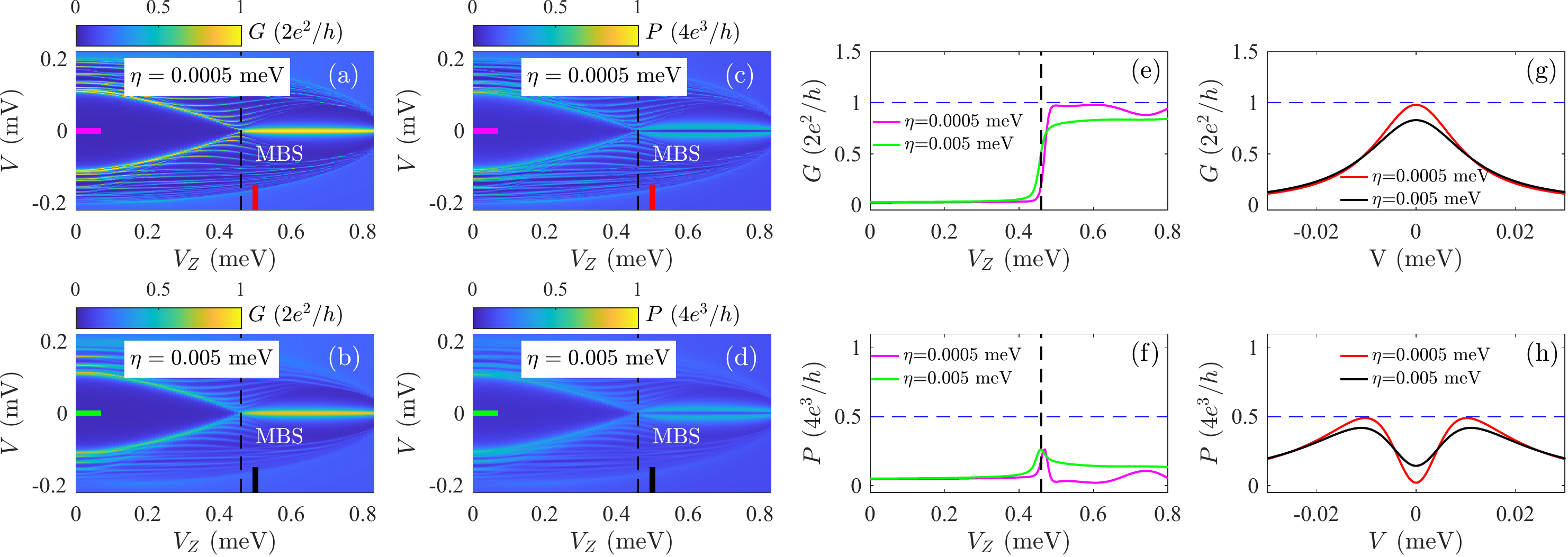}
\caption{Simulations of wire $\#1$ with a negligible ($\eta=0.0005$ meV) or considerable ($\eta=0.005$ meV) dissipation. The black vertical dashed lines in (a)–(f) mark the critical $V_Z$ at which the topological superconducting phase transition occurs.}\label{Fig_MBS_dissipation}
\end{figure*}

\begin{figure*}[htbp]
\centering
\includegraphics[width=\columnwidth]{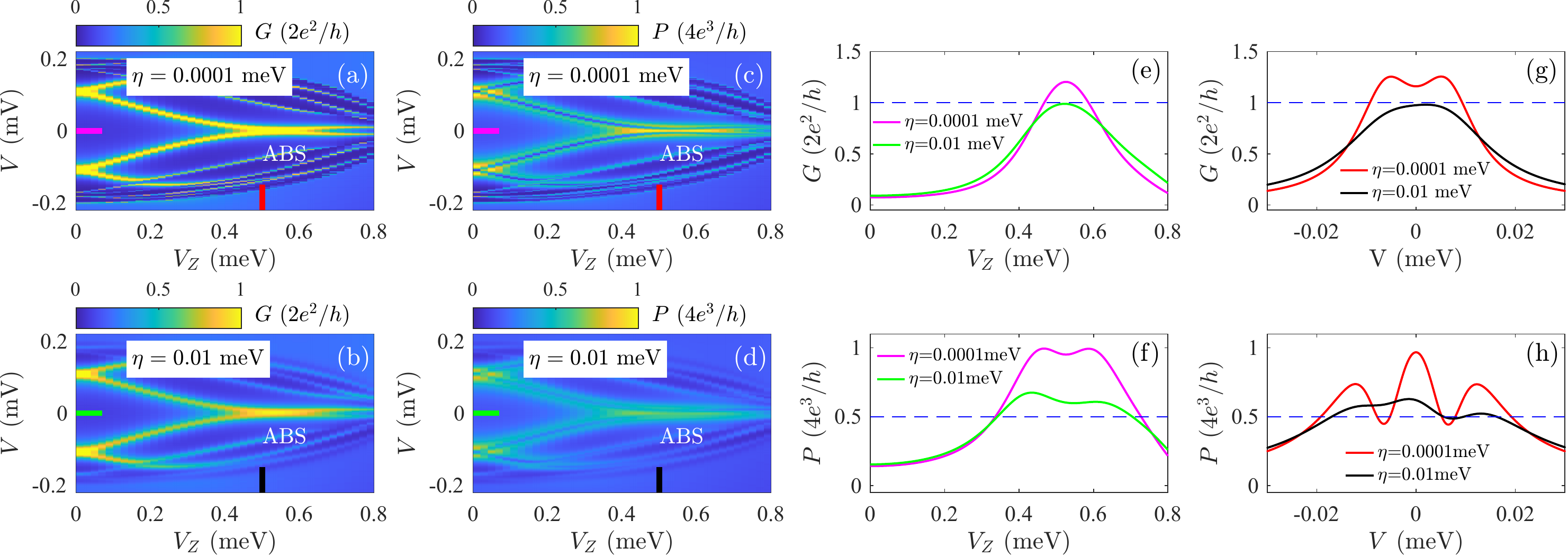}
\caption{Simulations of a disordered single-subband hybrid nanowire (referred to as wire $\#3$) with a negligible ($\eta=0.0001$ meV) or considerable ($\eta=0.01$ meV) dissipation. The nearly quantized conductance plateau in (b) results from a near-zero-energy ABS suffering from dissipation, as clearly shown by the line cuts in (e) and (g).}\label{Fig_ABS_dissipation}
\end{figure*}

\begin{figure*}[htbp]
\centering
\includegraphics[width=\columnwidth]{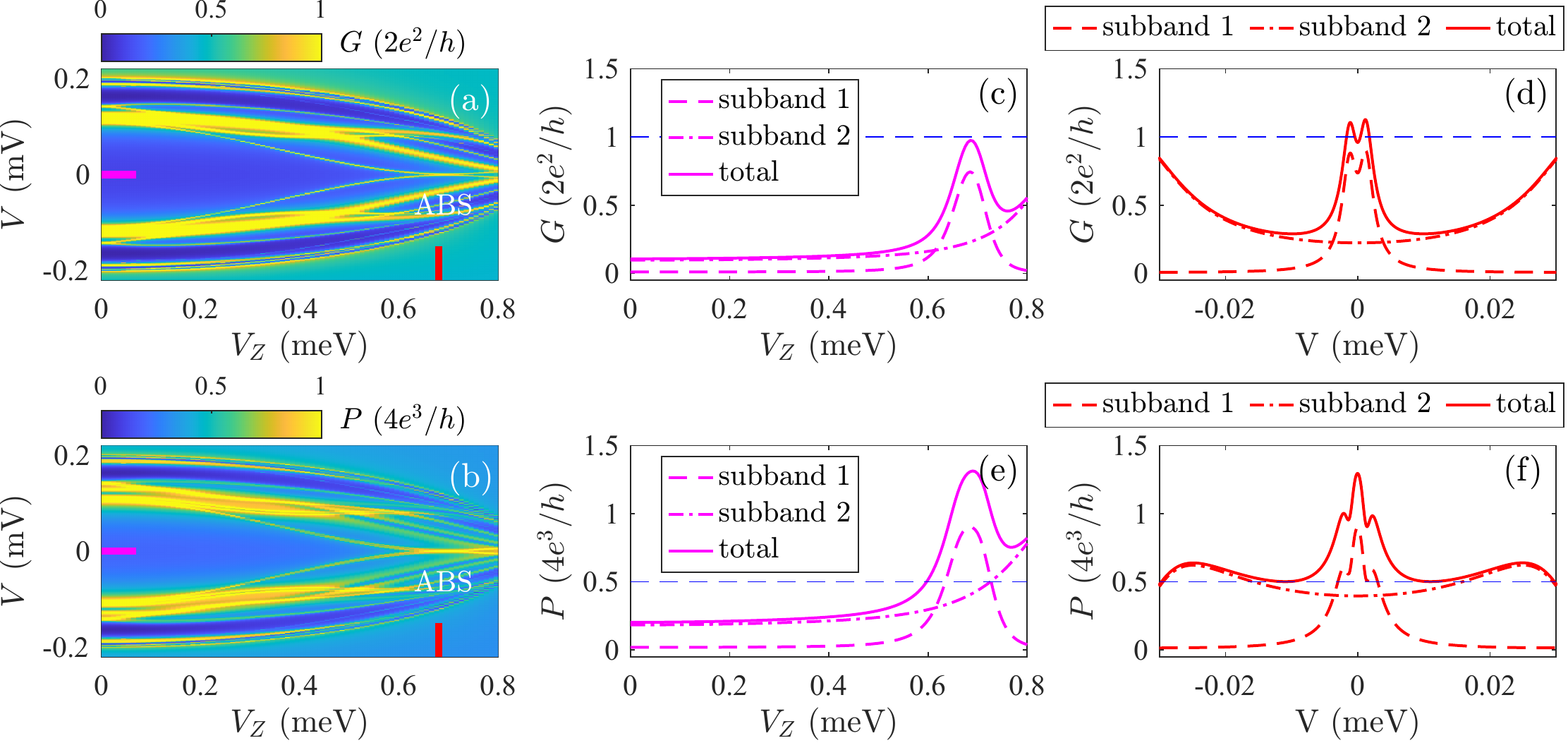}
\caption{Simulations of a disordered two-subband hybrid nanowire (referred to as wire $\#4$). In (c)–(f), the contributions from each subband and the total are shown. The nearly quantized conductance plateau in (a) results from an ABS in subband 1 and the background conductance of subband 2, as clearly shown by the line cuts in (c) and (d).}\label{Fig_ABS_and_background}
\end{figure*}

\begin{figure*}[htbp]
\centering
\includegraphics[width=\columnwidth]{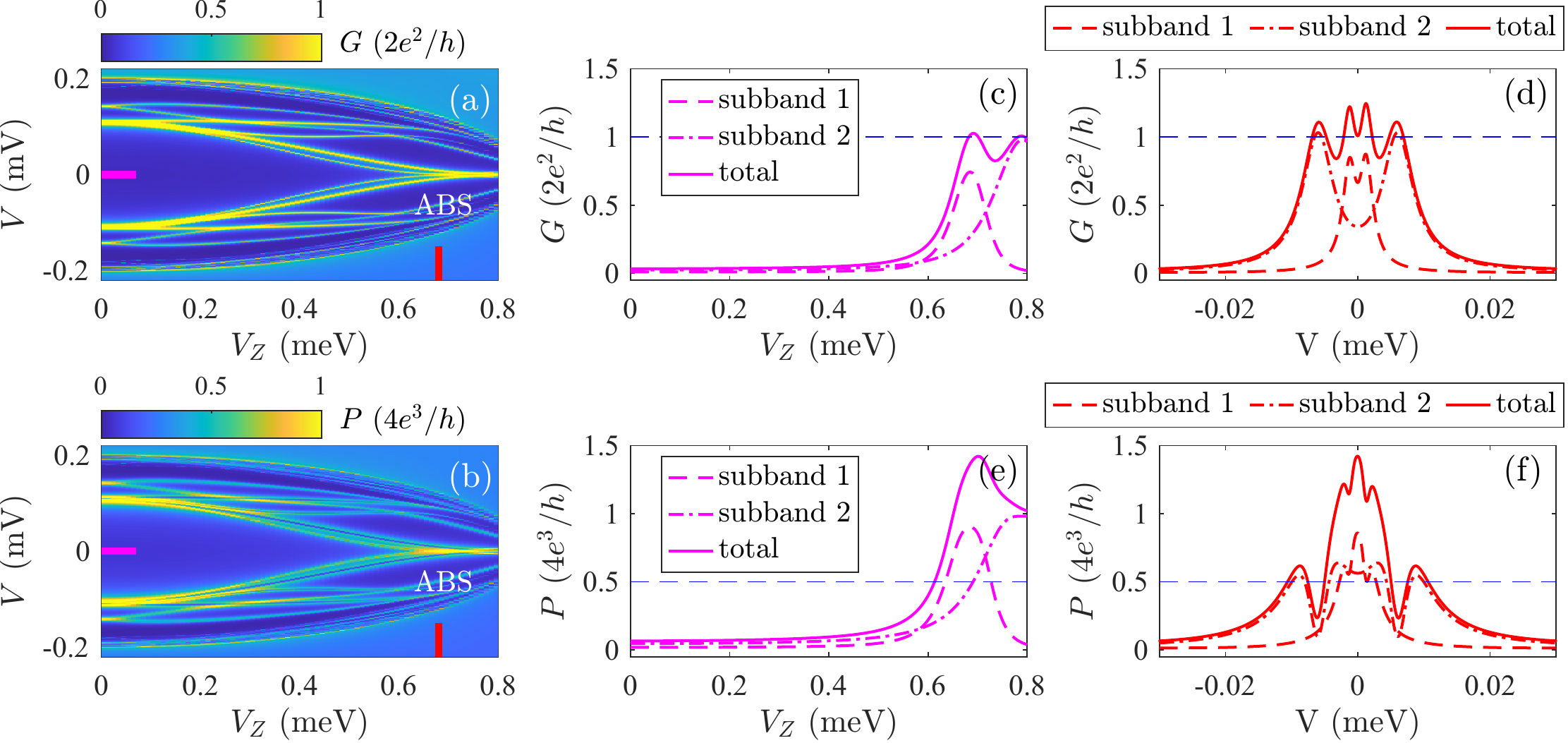}
\caption{Simulations of another disordered two-subband hybrid nanowire (referred to as wire $\#5$). The nearly quantized conductance plateau in (a) results from two low-energy ABSs in subbands 1 and 2, respectively, as clearly shown by the line cuts in (c) and (d).}\label{Fig_ABS_and_ABS}
\end{figure*}

In Figs.~\ref{Fig_DOS}(a) and \ref{Fig_DOS}(b), we show the maps of the local density of states $\rho(x,\varepsilon)$ of the two low-energy bound states indicated by the colored bars in Figs.~5(a) and 5(b), respectively, in the main text. As we can see, both bound states have near-zero energies. However, the one in Fig.~\ref{Fig_DOS}(a) is localized at the two ends of the nanowire, while the one in Fig.~\ref{Fig_DOS}(b) is extended over the whole nanowire. Therefore, they are MBS and near-zero-energy ABS, respectively.

The simulations in Fig.~5 in the main text are performed with negligible dissipation on hybrid nanowires. However, in certain experimental situations, a hybrid nanowire may couple to its surrounding environment, resulting in dissipation effects on superconducting bound states. In Figs.~\ref{Fig_MBS_dissipation}(a)--\ref{Fig_MBS_dissipation}(d), we illustrate the impact of dissipation (parameterized by $\eta$) on the Majorana nanowire, i.e., wire $\#1$. The line cuts in Figs.~\ref{Fig_MBS_dissipation}(e) and \ref{Fig_MBS_dissipation}(g) demonstrate that dissipation suppresses the quantized conductance plateau or ZBCP induced by an MBS to below $2e^2/h$, which is consistent with the findings of Ref.~\cite{liu2017role}. However, as shown in Fig.~\ref{Fig_MBS_dissipation}(h), the line cuts reveal that a broad zero-bias dip in $P(V)$ and $P_{max}\le 2e^3/h$ still remain, unaffected by dissipation. Moving on to Figs.~\ref{Fig_ABS_dissipation}(a)--\ref{Fig_ABS_dissipation}(d), we present maps of $G$ and $P$ for another disordered single-subband hybrid nanowire (referred to as wire $\#3$) under two dissipation scenarios: negligible ($\eta=0.0001$ meV) and considerable ($\eta=0.01$ meV) dissipation. The line cuts in Figs.~\ref{Fig_ABS_dissipation}(e) and \ref{Fig_ABS_dissipation}(g) explicitly demonstrate that a near-zero-energy ABS induced conductance plateau or ZBCP above $2e^2/h$ can be suppressed by dissipation to near $2e^2/h$, resembling the conductance behavior of a single MBS/QMBS. However, the line cuts in Figs.~\ref{Fig_ABS_dissipation}(f) and \ref{Fig_ABS_dissipation}(h) reveal that the associated $P(V)$ can exceed $2e^3/h$, indicating that the nearly quantized conductance plateau or ZBCP is associated with ABSs.

In Fig.~\ref{Fig_ABS_and_background}, we perform simulations of a disordered two-subband hybrid nanowire (referred to as wire $\#4$) to explore a possible scenario in which a nearly quantized conductance plateau consists of an unquantized conductance plateau (associated with a near-zero-energy ABS in subband 1) and a smooth background conductance (associated with the quasi-particle continuums in subband 2), as illustrated in Fig.~\ref{Fig_ABS_and_background}(c). Notably, the total $P(V)$ in Fig.~\ref{Fig_ABS_and_background}(f) exhibits a zero-bias peak and exceeds $2e^3/h$ over a significant bias range, indicating that the nearly quantized conductance plateau is associated with ABSs. 

\begin{table}[b!]
\centering
\tabcolsep=0.6cm
\caption{Results obtained from the effective model calculations of a probe coupled to different bound states. All cases exhibit a nearly quantized ZBCP in the $G(V)$ curves.}\label{tb3}
\begin{tabular}{ccccc}
\hline
\hline
Cases &\makecell[c]{a broad zero-bias dip \\ in $P(V)$} &\makecell[c]{a double-peak in $P(V)$ \\ at $eV\approx \pm3k_B T$} &\makecell[c]{$P_{max} >2e^3/h$}\\
\hline
\makecell[c]{An MBS/QMBS, \\ see Fig.~3 and Fig.~\ref{Fig_variousToneboundstate}} &\checked &$\times$   &$\times$\\
\hline
\makecell[c]{An ABS, \\ see Fig.~3 and Fig.~\ref{Fig_variousToneboundstate}}  &$\times$ &\checked &\checked\\
\hline
\makecell[c]{An MBS/QMBS and an ABS, \\ see Fig.~4}  &\checked&$\times$ &\checked\\
\hline
\makecell[c]{Two ABSs, \\ see Fig.~4}  &$\times$ &\checked  &\checked\\
\hline
\hline
\end{tabular}
\end{table}

\begin{table}[htbp]
\centering
\tabcolsep=0.6cm
\caption{Zero-temperature simulations of semiconductor-superconductor hybrid nanowires exhibiting nearly quantized conductance plateaus.}\label{tb4}
\begin{tabular}{cccc}
\hline
\hline
Cases &\makecell[c]{a broad zero-bias dip \\in $P(V)$} &\makecell[c]{a zero-bias peak \\in $P(V)$} &\makecell[c]{$P_{max} >2e^3/h$}\\
\hline
\makecell[c]{An MBS in a clean single-subband \\ nanowire, see Fig.~5}  &\checked &$\times$ &$\times$\\
\hline
\makecell[c]{An MBS in a clean single-subband \\ nanowire with dissipation, see Fig.~\ref{Fig_MBS_dissipation}}  &\checked &$\times$ &$\times$\\
\hline
\makecell[c]{An ABS in a disordered  
single-subband \\ nanowire, see Fig.~5}  &$\times$  &\checked &\checked\\
\hline
\makecell[c]{An ABS in a disordered single-subband \\nanowire with dissipation, see Fig.~\ref{Fig_ABS_dissipation}}  &$\times$  &\checked &\checked\\
\hline
\makecell[c]{An ABS in a disordered two-subband \\ nanowire, see Fig.~\ref{Fig_ABS_and_background}}  &$\times$  &\checked &\checked\\
\hline
\makecell[c]{Two ABSs in a disordered two-subband\\ nanowire, see Fig.~\ref{Fig_ABS_and_ABS}}  &$\times$  &\checked &\checked\\
\hline
\hline
\end{tabular}
\end{table} 

In Fig.~\ref{Fig_ABS_and_ABS}, we simulate another disordered two-subband hybrid nanowire (referred to as wire $\# 5$) and observe a nearly quantized conductance plateau, see Fig.~\ref{Fig_ABS_and_ABS}(a). The line cuts in Figs.~\ref{Fig_ABS_and_ABS}(c) and \ref{Fig_ABS_and_ABS}(d) reveal that the conductance plateau arises from two low-energy ABSs in different subbands. Furthermore, the associated $P(V)$ curve in Fig.~\ref{Fig_ABS_and_ABS}(f) exhibits a zero-bias peak and significantly exceeds $2e^3/h$ around zero-bias, indicating that the nearly quantized conductance plateau is associated with ABSs.

For clarity, we summarize the results obtained from the effective model calculations and numerical simulations in Tables \ref{tb3} and \ref{tb4}, respectively.

%\bibliographystyle{apsrev4-1-title}
%\bibliographystyle{apsrev4-1}
%\bibliography{refs-Majorana}
